\documentclass[12pt,preprint]{aastex}

\slugcomment{Revised version: \today}

\shorttitle{Young Earths}
\shortauthors{Reid et al}

\begin{document}

\def\pedant{Mart{\'{\i}}n }
\def\etal{{\sl et al.}}
\def\etall{{\sl et al. }}
\def\pma{$\arcsec$~yr$^{-1}$ }
\def\kms{km~s$^{-1}$ }
\def\msun{$M_{\odot}$}
\def\rsun{$R_{\odot}$}
\def\lsun{$L_{\odot}$}
\def\halpha{H$\alpha$}
\def\hbeta{H$\beta$}
\def\hgama{H$\gamma$}
\def\hdelta{H$\delta$}
\def\Teff{T$_{eff}$}
\def\logg{$log_g$} 

\title {Searching for Earth analogues around the nearest stars: the disk age-metallicity relation and the age distribution in the Solar Neighbourhood}

\author{I. Neill Reid}
\affil{Space Telescope Science Institute, 3700 San Martin Drive, Baltimore,
MD 21218, USA}
\email{inr@stsci.edu}

\author{Edwin L. Turner}
\affil{Princeton University Observatory, Peyton Hall, Princeton, NJ 08544, USA}
\email{elt@astro.princeton.edu }

\author{Margaret C. Turnbull, M. Mountain, Jeff A. Valenti}
\affil{Space Telescope Science Institute, 3700 San Martin Drive, Baltimore,
MD 21218, USA}

\clearpage

\begin {abstract}
The chemical composition of Earth's atmosphere has undergone substantial evolution over the course of its history. It is possible, even likely, that terrestrial planets in other planetary systems have undergone similar changes; consequently, the age distribution of nearby stars is an important consideration in designing surveys for Earth-analogues. Valenti \& Fischer (2005) provide age and metallicity estimates for 1039 FGK dwarfs in the Solar Neighbourhood. Using the Hipparcos catalogue as a reference to calibrate potential biases, we have extracted volume-limited samples of nearby stars from the Valenti-Fischer dataset. Unlike other recent investigations, our analysis shows clear evidence for an age-metallicity relation in the local disk, albeit with substantial dispersion at any epoch. The mean metallicity increases from $\sim$-0.3 dex at a lookback time of $\sim$10 Gyrs to $\sim$+0.15 dex at the present day. Supplementing the Valenti-Fischer measurements with literature metallicity data to give a complete volume-limited sample, the age distribution of nearby FGK dwarfs is broadly consistent with a uniform star-formation rate over the history of the Galactic disk. In striking contrast, most stars known to have (gas giant) planetary companions are younger than 5 Gyrs; however, stars with planetary companions within 0.4 AU have a significantly flatter age distribution, indicating that those systems are stable on timescales of many Gyrs. Several of the older, lower metallicity host stars have enhanced [$\alpha$/Fe] ratios, implying membership of the thick disk. If the frequency of terrestrial planets is also correlated with stellar metallicity, then the median age of such planetary system is likely to be $\sim$3 Gyrs. We discuss the implications of this hypothesis in designing searches for Earth analogues among the nearby stars.

\end{abstract}


\section {Introduction}

The main goal of the NASA Navigator program is the discovery and characterisation of Earth-like planets in orbit around stars other than the Sun. Conceptually, this quest can be considered as two separate tasks: first, imaging a terrestrial-mass planet within the habitable zone of a star  (or stars) in the immediate Solar Neighbourhood; second, determining whether that planet is, was, or might in the future be capable of supporting life. The first task poses some considerable technical challenges, and NASA is currently investigating two different architectures, optical coronagraphy and near-infrared interferometry, that might be employed in a future Terrestrial Planet-Finder mission or missions. The second task requires a thorough understanding not only of key spectral signatures in Earth's atmosphere at the present day, but also of how those signatures have evolved over time. 

Given the technical constraints, observational surveys for terrestrial planets must focus on the stars nearest the Sun, particularly solar-type stars within 20-30 parsecs. The characteristics of those stars, notably the age distribution, are likely to influence strongly the characteristics of the initial set of planet detections. In addition, the current roster of gas-giant extrasolar planets exhibits a strong correlation between planetary frequency and metallicity (Gonzalez, 1998, 2000; Santos et al, 2000, 2004; Reid, 2002; Fischer \& Valenti, 2005). Recent observations suggest that this trend may weaken for Neptunian-mass giant planets (Udry et al, 2006), However, if the correlation extends to terrestrial planets, then any underlying correlation between age, $\tau$, and metallicity will affect the likely age distribution of nearby planetary systems.

Considerable effort has been devoted to modeling the evolution of Earth's atmosphere over the $\sim4.5$ Gyrs history since its formation. As reviewed by Kasting \& Catling (2003), those models indicate that the atmosphere was CO$_2$-rich (mixing ratio $\sim10\%$) during the initial $\sim500$ Myrs of the Hadean phase. The CO$_2$ fraction decreased (by a factor 10) over the first $\sim200$ Myrs of the Archaean period (which spans, in total, lookback times from $\sim3.8$ to 2.5 Gyrs); over the same period, the atmospheric methane content rose to $\sim1\%$, probably driven by archaeal and bacterial methanogenesis. This probably resulted in a significant increase in the greenhouse effect (Pavlov, Brown \& Kasting, 2001). Earth's atmosphere remained largely anoxic until the early Proterozoic, $\sim2.3$ Gyrs ago, when the O$_2$ fraction started to rise slowly, probably due to oxygenic photosynthesis by terrestrial organisms (Kasting \& Catling, 2003). The oxygen content remained below 1\% until $\sim1$ Gyrs ago, and did not reach current levels until the mid-Phanerozoic, $\sim 0.5$ Gyrs ago. This change probably coincided with the first appearance of extensive surface vegetation during the Devonian period (Berner, 1997; Beerling \& Royer, 2002).

The spectroscopic characteristics of Earth's atmosphere underwent substantial evolution as a result of these atmospheric compositional changes. Kaltenegger, Traub \& Jucks (2006) have recently modeled the expected spectral evolution at optical/near-infrared and mid-infrared wavelengths. As one might expect, the rise in oxygen leads to significant changes, notably stronger H$_2$O absorption at near- and mid-infrared wavelengths and the appearance of O$_2$ and ozone absorption in the optical. With the absence of extensive surface vegetation during the first $\sim4$ Gyrs, the `red edge' vegetative biosignature at $\sim7200\AA$ (Seagar et al, 2005) has been detectable for no more than the last 500 Myrs or so of Earth's history. Large microbial mats were probably present on Earth's oceans at earlier epochs, but sea water's high opacity is likely to mask and weaken any associated spectroscopic features (Knacke, 2003).

Of course, extrapolating Earth's evolutionary sequence to other terrestrial planets carries the caveat that the relative timescales are based on the statistics of one. Earth may prove to be as useful a paradigm for atmospheric evolution as the Solar System has been for the orbital distribution of giant planets\footnote{We have no evidence for atmospheric evolution beyond an age $\tau \sim4.5$ Gyrs; however, (post-)civilised planets may well prove to be characterised by increased abundances of greenhouse gases, rather than post-atomic debris.}. Nonetheless, a well-balanced search for nearby habitable planets should take into account the potential distribution of atmospheric composition.

The present paper examines the age distribution of nearby stars, and the consequent likely age distribution of nearby planetary systems. The following section outlines the selection of a suitably representative, and well-studied, sample of nearby stars; \S3 considers the insight those stars afford into the age-metallicity relation for the Galactic disk, and the age distribution of local main sequence stars; and \S4 discusses these results in the context of terrestrial planetary systems; and \S5 presents our conclusions.

\section {A local reference sample}

Investigating the underlying properties of nearby solar-type stars requires a sample of well-studied solar-type stars with reliable age and metallicity measurements, that are representative of the local stellar populations. There are a only a limited number of potential sources for a reference sample, and due care needs to be exercised in selecting a reliable dataset.

Nordstr\"om et al (2004, N04) have catalogued the intrinsic parameters of $\sim14,000$ FGK stars. The dataset includes a near-complete sample of solar-type stars within 40 parsecs of the Sun, although only a subset has age estimates. The metallicities are derived from Str\"omgren {\sl uvby} photometry, where the calibration follows the Sch\"uster \& Nissen (1989) formulation and is based on 72 dwarfs with high-resolution spectroscopic abundance analyses. Photometric metallicity calibrations can be problematic, as discussed by Reid (2002). Haywood (2002) has derived an alternative calibration for {sl uvby} photometry, using the same formalism as Sch\"uster \& Nissen\footnote {Haywood has also demonstrated that the original calibration by Sch\"uster \& Nissen (1989) is subject to systematic errors, mainly due to incorrect placement of the Hyades reference sequence.}. The uppermost panel in Figure 1 compares the N04 metallicities against Haywood's revised calibration; it is clear that there are systematic differences between the two analyses. Haywood (2002) has shown that his calibration is in good agreement with spectroscopic datasets. Since the NO4 age estimates are derived from isochrone fitting, systematic errors in the metallicities translate to systematic offsets in age.

An alternative reference dataset is provided by Valenti \& Fischer's (2005; VF05) detailed analysis of nearby solar-type stars. The study is based on high-resolution echelle spectra obtained in the course of the Keck, Lick and AAT planet search programs. The VF05 catalogue includes Fe, Na, Si, Ti and Ni abundances, with precision $\pm0.03$ dex; gravities to $\pm$0.06 dex; projected rotational velocities to $\pm$0.5 \kms; temperatures to $\pm44$K; and luminosities with an average uncertainty of $\pm6\%$. The two lower panels in Figure 1 compare the VF05 abundance measurements against the N04 and Haywood metallicity calibrations for stars with {\sl uvby} data. Table 1 lists the mean offsets, and the rms dispersion about the mean, as a function of $(b-y)$. While the dispersions are similar, there is clearly better agreement between the VF05 and Haywood metallicity scales.

Valenti \& Fischer estimate ages for stars in their dataset by matching the empirical (Log(L), log(T$_e$)) data against the Yonsei-Yale isochrones (Demarque et al, 2004), interpolating in metallicity as necessary. They derive a probability distribution for each star, taking account different potential evolutionary states (main sequence, subgiant or giant branch), tabulating the median value and the $\pm1\sigma$ uncertainties for each star\footnote{ We note that some of the averaged stellar parameters tabulated by Valenti \& Fischer in the published paper were actually determined using a slightly different method than outlined in the text of that paper. We have used revised values that were calculated following the exact procedures described in the paper.}. 

The VF05 catalogue therefore provides reliable estimates of intrinsic stellar quantities, but the sample is not complete in either a volume-limited or magnitude-limited sense. It is therefore crucial to compare the overall properties, particularly the abundance distribution, against a representative sample of disk stars before identifying a suitable subset for probing the age distribution of local stars. This process is described in the following sections.

\subsection {The characteristics of the VF05 dataset}

The stars in the VF05 dataset constitute the parent sample for the radial-velocity planet search programs. The sample is selected based primarily on colour and apparent magnitude, with an {\sl a priori} bias against known close binary stars. There is an {\sl a posteriori} selection for chromospherically quiescent stars:  stars that show significant chromospheric emission are flagged as likely to exhibit substantial velocity jitter, and are not targeted for follow-up radial-velocity monitoring. The exclusion of chromospherically active stars leads to a bias against young stellar systems in the planet search program. However, it is important to emphasise that this age bias is {\sl not} present in the parent VF05 dataset.

Figure 2 plots the (M$_V$, distance) distribution and (M$_V$, (B-V)) colour-magnitude diagram of the 1039 stars in the VF05 sample. Since most observations were obtained from either the Lick or Keck Observatories, the sample lies predominantly at northern declinations. By design, the sample consists primarily of main-sequence F, G and K stars, with the majority having apparent magnitudes brighter than V=8. The latter criterion accounts for the decrease in the effective distance limit with decreasing luminosity. Most stars lie within 40 parsecs of the Sun, with a tail of higher luminosity F and early-type G stars extending beyond 80 parsecs. 

Making due exception for the youngest members, which may retain kinematic signatures of their formation regions, the Galactic disk is generally regarded as a well-mixed stellar population, both kinematically and spatially. Under those circumstances, the optimum method of sampling the population characteristics is the construction of a complete, volume-limited sample. While we cannot directly apply this selection criterion to the incomplete VF05 dataset, we can match that catalogue against a complete, volume-limited sample from another source, the Hipparcos catalogue (ESA, 1997), and define a volume-limited subset that offers the best prospects of a representative sampling of local stars. 

The Hipparcos astrometric satellite obtained milliarcsecond precision astrometry of approximately 118,000 stars, including most stars brighter than V=9. As a result, the Hipparcos catalogue includes the overwhelming majority of solar-type (F, G, early-K) stars within 40 parsecs of the Sun. All save nine of the stars in the VF05 sample are also in the Hipparcos catalogue. Consequently, we can determine the fractional completeness of the VF05 sample as a function of distance. Figure 3 plots those data, segregated by absolute magnitude. Clearly, the completeness drops significantly at distances beyond 25 to 30 parsecs and at absolute magnitudes fainter than M$_V \sim 6.5$. This needs to be taken into account in defining the appropriate selection criteria, as described in the following section.

\subsection{A representative sample of local disk stars}

Taking the VF05 dataset as a reference, we need to identify a subset of these stars that provides representative sampling of the local population, integrated over formation history of the disk. Luminous main-sequence stars have higher mass, and therefore shorter lifetimes; consequently, a stellar sample that includes high-luminosity stars preferentially samples recent star formation epochs. This argues against including stars with main-sequence lifetimes that are substantially less than age of the Galactic disk ($\sim9$ Gyrs). We therefore exclude stars with M$_V < +4$. As an apparent-magnitude limited sample, the distance limit shrinks at fainter absolute magnitudes, and the sharp decline in completeness at M$_V > 6.5$ effectively sets low luminosity limit.

Combining the absolute magnitude constraints with the spectral type/colour selection criteria and the distance distributions shown in Figures 2 and 3, we have defined two volume-limited samples of nearby FGK stars. Both are drawn from the Hipparcos catalogue, and in each case the subset of VF05 stars is sufficiently large that it is likely to be representative of the local disk. The sub-samples are not independent, but comparing results derived from these datasets allows some assessment of the potential for systematic biases. 

The samples are defined as follows:
\begin{description}
\item[Sample A:] stars with absolute magnitudes $4 < M_v < 6$ and distances $d < 30$ pc; this sample consists of 565 stars, including 297 from the VF05 dataset. These are $\sim$F5 to K0 dwarfs, with masses in the range $\sim1.25$ to $\sim0.85 M_\odot$.  
\item[Sample B:] stars with absolute magnitudes $4 < M_v < 6.5$ and distances $d < 25$ pc; this sample consists of 409 stars, including 239 from the VF05 dataset. These are $\sim$F5 to K4 dwarfs, with masses in the range $\sim1.25$ to $\sim0.8 M_\odot$.  
\end{description}
There is substantial overlap between the two samples, with 183 VF05 stars and 133 non-VF05 stars in common. 

Our goal is to use the ages estimated for the VF05 stars in these two samples to probe the age-distribution (and age-metallicity relation) of the local disk. It is therefore important to compare the relative properties of the VF05 and non-VF05 stars in Samples A and B. Figures 4 and 5 compare the (M$_V$, (B-V)) colour-magnitude distributions. In both cases, the VF05 dataset shows less scatter above the main sequence, reflecting the explicit omission of known close binaries from planet search programs. With this exception, the colour-magnitude distributions are very similar. 

Next to mass and age, metallicity is the most important factor in determining the intrinsic properties of a star; moreover, metallicity appears to play a key role in determining the likelihood of forming a planetary system. It is therefore important to gauge whether the VF05 stars in Samples A and B are representative of the underlying abundance distribution of disk stars. To carry out this comparison, we need to compare like with like. All the VF05 stars have accurate high-resolution spectral analyses, but such data do not exist for most of the non-VF05 stars. Therefore, for this comparison, we must turn to lower-accuracy abundance estimators, specifically Str\"omgren photometry.

Abundances estimates for the VF05 and non-VF05 stars in Samples A and B were compiled as follow:
\begin{itemize}
\item All stars in Samples A and B have been cross-referenced against the Hauck \& Mermilliod (1998) {\sl uvby} catalogue. In Sample A, 259 VF05 (87\%) and 234 non-VF05 (87\%) stars have {\sl uvby} data; in Sample B, 188 VF05 stars (79\%) and 137 non-VF05 stars (80\%) have {\sl uvby} data. We have used the Haywood (2002) calibration to estimate abundances for al these stars.

\item In the case of VF05 stars that lack Str\"omgren data, we adopt the [Fe/H] values determined by Valenti \& Fischer (2005).  Figure 1 shows that the zeropoint of these measurements is consistent with the Haywood calibration.
 
\item For Hipparcos stars that are not in the VF05 dataset, we have searched the literature for alternative metallicity measurements. We have located data for a further 31 stars from Sample A and 24 additional stars from Sample B. Those measurements are taken from several sources, including Kotoneva et al (2002 - 23 stars; photometric indices), Zakhozhaj \& Shaparenko(1996 - 5 stars; UBV indices), Santos et al (2002 - 3 stars; high resolution spectroscopy), Carney et al (1994 - 2 stars; UBV indices), Fuhrmann (2004 - 2 stars; high resolution spectroscopy) and Soubiran \& Girard (2005 - 1 star; high resolution spectroscopy). 

\end{itemize}
Combining data from all these sources leaves only 11 stars in Sample A and 10 stars in Sample B lacking metallicity estimates.

Figure 6 compares the abundance distributions derived for the VF05 and non-VF05 subsets from Samples A and B. The left hand panels plot binned differential distributions, while the right hand panels show the cumulative distributions together with the fractional completeness of the VF05 dataset as a function of [Fe/H]. It is clear that both VF05 sub-samples include a higher proportion of metal-rich stars than an unbiased sampling of the local disk population. Specifically, in Sample A, 47\% of the VF05 stars have [Fe/H]$>$0.1, while only 27\% of the non-VF05 stars are this metal rich; similarly, 40\% of the VF05 stars in Sample B meet this metallicity threshold, but only 26\% of the non-VF05 stars. The origin of this bias is not clear, but may partly reflect that fact that the VF05 sample is magnitude-limited and luminosity increases with metallicity for main sequence stars. In addition, there may be a tendency to include metal-rich stars in the planet-search survey. This bias must be taken into account when using the VF05 dataset to infer the likely age distribution of local disk stars, as discussed further in the following section.

\subsection {Summary}

Our reference datasets for the local disk population are two volume-limited samples of FGK dwarfs drawn from the Hipparcos catalogue. There is significant overlap between the two samples. In both cases, over half the stars are included in the Valenti \& Fischer analysis, and have reliable spectroscopic metallicity estimates and isochrone-based age estimates. We use the VF05 data for these stars as the principal guide to the age-metallicity relation derived in the following section. Almost all of the remaining stars in the two datasets, Samples A and B, have photometrically-based metallicity estimates, primarily derived from Haywood's calibration of Str\"omgren photometry; these stars, however, lack direct age estimates. 

\section{The characteristics of the local disk population}

The principal aims of the current investigation are a determination of the age distribution of local disk stars, and hence an estimate of the likely age distribution of nearby planetary systems. 

\subsection {The age-metallicity relation for the Galactic disk}

The age-metallicity relation (AMR) is a fundamental to understanding Galactic evolution, providing an empirical measure of how star formation has enriched the interstellar medium (ISM) over the history of the disk. The form, indeed the existence, of an AMR has been debated in the astronomical literature for well over thirty years. Theoretical models lead to an expectation of increasing metallicity with time, as succeeding generations of stars return nucleosynthetic products to the 
ISM, and initial analyses (e.g. Twarog, 1980) were consistent with these expectations. However, the existence of old, metal-rich clusters, such as NGC 6791, M67 and NGC 188, indicates significant dispersion in metallicity at lookback times $t_L > 5$ Gyrs. Moreover, in a highly influential paper, Edvardson et al (1993, E93) analysed high-resolution spectroscopic data for 189 field F and G dwarfs, deriving ages by matching star to theoretical isochrones in the ($\log{T_{eff}}, \log{L}$) plane. E93 interpreted their results as showing a weak AMR, with a substantial dispersion in metallicity, $\sigma_{[Fe/H]} \sim 0.25$dex, for $t_L < 10$ Gyrs. 

Several subsequent investigations, including Feltzing et al (2001), Ibukiyama \& Arimoto (2002) and N04, arrived similar conclusions to E93, finding little evidence for systematic variations in $\langle[Fe/H]\rangle$ with age, and a consistently broad dispersion. All of these studies are based on photometric data, notably Str\"omgren photometry. In contrast, Rocha-Pinto et al (2000, RP00) have used chromospheric activity to estimate ages for 525 nearby main-sequence dwarfs; they find a significant trend in mean abundance, with significant lower dispersion ($\sigma_{[Fe/H]} \sim 0.12$ dex) at a given age. Recently, Pont \& Eyer (2004) have shown that significant bias can be present in the traditional isochrone-matching technique, with a tendency to favour ages close to the main-sequence lifetime for individual dwarfs. They have re-analysed the E93 sample using Bayesian techniques, and find results closer to the Rocha-Pinto et al analysis, with a significant trend in mean metallicity with age and a dispersion  $\sigma_{[Fe/H]} < 0.15$ dex at a given age.

Figure 7 plots results from the E93, RP00 and N04 analyses together with the AMR defined by VF05 stars in Samples A and B. Ages and metallicities for the latter stars are taken from the Valenti \& Fischer (2005) analysis (revised from the published values as described in footnote \#3), and we plot formal uncertainties for a representative subset in Figure 7. Valenti \& Fischer present an extensive comparison between the results from their analysis and literature data, showing good agreement in T$_{eff}$ and [Fe/H] (their Figure 18 and 19). In particular, there are 26 stars in common between the VF05 and E93 datasets, and Figure 8 compares metallicities and ages for those stars. There is a $\sim0.1$ dex offset in [Fe/H], with the VF05 data systematically more metal-rich; this is consistent with the literature data comparisons made by Valenti \& Fischer. On the other hand, there are 570 stars in common between the VF05 and N04 datasets, of which 386 have age estimates. Figure 8 shows that there is essentially no correlation between the ages derived in these two analyses.

All datasets plotted in Figure 7 include stars with formal ages that exceed the WMAP estimate of 13.7 Gyrs for the age of the Universe under a $\Lambda$-CDM cosmology (Verde \etal, 2003). While the stellar evolutionary timescale is defined independently of cosmological considerations, there is broad concensus between the ages estimated for halo globular clusters (e.g. Chaboyer \etal, 1998) and the WMAP result. However, determining reliable ages for individual stars is more problematic, particularly for older stars, which are more susceptible to the isochrone-matching bias identified by Pont \& Eyer (2004). Age estimates are more accurate at $\tau < 5$ Gyrs; for example, the VF05 analysis includes the Sun, which is assigned an age of $4.3^{+1.7}_{-1.6}$ Gyrs. Overall, it is likely that the age rankings in the VF05 sample are more reliable than the absolute age estimates.

It is clear that the age-metallicity distribution of the VF05 data exhibits a more consistent variation in mean abundance with time than either the E93 or N04 datasets, and a shallower gradient (at least over the last 2-3 Gyrs) than the RP00 analysis. Note that the Sun has average properties for its age in the VF05 analysis, whereas it is abnormally metal-rich when matched against the RP00 AMR. Fitting a linear relation to the data in Sample A gives,
\begin{equation}
[Fe/H] = (0.177\pm0.020) \ - \ (0.040\pm0.003) \times t_L, \quad \sigma = 0.18 {\rm dex}
\end{equation}
where $t_L$ is lookback time in Gyrs. Note that, since [Fe/H] is a logarithmic quantity, a linear relation formally implies that either the star formation rate or the yield (or both) increased with time over the history of the Galactic disk. Fitting a second-order polynomial gives
\begin{equation}
[Fe/H] = 0.118 \ - \ 0.0139 t_L \ - \ 0.00197 t_L^2, \quad \sigma=0.18 {\rm dex}
\end{equation}
Mean relations derived from the VF05 stars from Sample B are statistically indistinguishable, which is not surprising given the substantial overlap between the two samples. The presence of an AMR has potentially significant implications for the age distribution of nearby planetary systems, as discussed further in \S4. 

\subsection {Selection effects and the AMR}

Can the correlation between age and metallicity that is evident in the VF05 data be attributed to a selection effect? A bias of this type might arise if, for example, Samples A and B included disproportionate numbers of metal-rich F stars, whose short main-sequence lifetimes might lead to an apparent age-metallicity trend. We believe that such biases are unlikely for two main reasons. First, the stars are selected based on absolute magnitude, M$_V$, rather than colour, reducing the potential for metallicity-based selection bias. Second, and more important, the sample is volume-limited, rather than magnitude-limited; consequently, the absolute magnitude distribution reflects the local luminosity function, and short-lived, high luminosity stars are minor constituents. The VF05 datasets, which encompass more than half the stars in Samples A and B, include a higher proportion of nearby metal-rich stars (see \S2.2 and Figure 6). However, there is no indication of a selection bias towards young/metal-rich and old/metal-poor stars that could produce the correlations present in Figure 7. 

Figure 9 quantifies the potential for sample bias. Taking Sample B a reference, we have superimposed theoretical predictions from the Yonsei-Yale models (Demarque et al, 2004) for a range of masses at ages 0.4, 1.0 and 5.0 Gyrs and for [Fe/H] +0.38, 0.04 and -0.43 dex (Z= 0.04, 0.02 and 0.007, and no $\alpha$-element enhancement). The figure shows that, for metal-rich stars, the absolute magnitude limits employed to define Sample B correspond to mass limits $\sim0.95 < {M \over M_\odot} < 1.3$ at Hyades-like ($\sim400$ Myr) ages, and $\sim0.9 < {M \over M_\odot} < 1.25$ at Sun-like ($\sim5$ Gyr) ages; in contrast, the mass limits are limits $\sim0.85 < {M \over M_\odot} < 1.2$ and $\sim0.8 < {M \over M_\odot} < 1.1$ for mildly metal-poor stars at similar ages. Reid, Gizis \& Hawley (2004) show that the stellar mass function for the local disk can be modeled as a composite power-law, $\Psi(M) \propto M^{-\alpha}$. Using their formalism, all of these mass limits encompass similar proportions of stars with the appropriate age and metallicity: approximately 9\% of local stars with masses between 0.1 and 3.0$M_\odot$. 

As a further check for possible bias, Figure 9 plots the (M$_V$, age) and (M$_V$, [Fe/H]) distributions for the VF05 stars in Sample B. We note that the most luminous stars are younger than $\sim7$ Gyrs, with the upper age limit increasing at fainter absolute magnitudes; this is expected, given the mass range of the sample. Stars with M$_V >5$ ($M < 0.9 M_\odot$) span the full age range of the disk. The more luminous stars also tend to be metal-rich. However, Figure 9 shows that the age/metallicity distribution of the long-lived M$_V>5$ stars is entirely consistent with the linear AMR (equation (1)) derived from the 297 VF05 stars in Sample A.

Based on this discussion, we see no evidence that the age-metallicity relation outlined by the VF05 stars in Samples A and B is due to a selection bias.

\subsection {The age distribution of local stars}

We are using the VF05 sample as a guide to the likely age distribution of nearby stars\footnote{The age distribution of local stars is not identical with the age distribution of stars in the Galactic disk. Disk heating leads to an increase in stellar velocities with time, and a consequent thickening of the density distribution perpendicular to the Plane (Wielen, 1977). Younger stars spend a larger fraction of their time near the Plane, and are therefore proportionately over-represented in local samples. However, this effect is most important for stars younger than $\sim1-2$ Gyrs, with empirical studies showing relatively little variation in the disk scaleheight at older ages. The local disk sample also probably includes interlopers that originated in either the inner or outer disk, but migrated outward or inward due to dynamical interactions. }. As discussed in \S2, besides estimating median ages, Valenti \& Fischer used isochrone fitting to derive probability distributions to represent the age of each star. We have combined the individual distributions to derive the likely age distribution of the VF05 dataset. However, the preponderance of metal-rich stars (relative to the local field) is likely to lead to moderately young ($\tau <3$ Gyrs) stars being over-represented in the VF05 dataset. 

To compensate for this potential bias, we have inverted the linear $\tau$/[Fe/H] relation derived in the previous section, and estimated ages for the non-VF05 stars in samples A and B that have metallicity measurements. This formalism leads to age estimates exceeding 15 Gyrs for stars with [Fe/H]$<-0.42$ dex, so the results can only be regarded as indicative. The resultant distributions are plotted in Figure 10, which shows both the summed probability distributions for the VF05 datasets (hatched distributions) and the best-estimate age distributions for the full samples. (To simplify matters, we use the median ages for the VF05 stars in computing the latter distributions). 

The age distributions derived for Samples A and B are similar, as expected given the overlap between the samples. Moreover, the age distributions of the VF05 datasets and the full samples are similar. The median age of the full dataset is older by $\sim0.5$ Gyrs for both samples - 4.7 Gyrs and 5.3 Gyrs for Sample A, and 5.3 Gyrs and 5.75 Gyrs for Sample B. That offset is also expected: ages for the non-VF05 sample are based on [Fe/H]; the non-VF05 stars have a larger proportion of metal-poor stars; therefore adding those stars to the VF05 sub-samples must increase the median age. 

The field-star age distributions plotted in Figure 10 extend beyond $t_L = 13$ Gyrs. As discussed in \S3.1, these apparently ancient (pre-primordial?) stars are likely to be a product of the inherent uncertainties (and biases) in age-dating; the true ages are likely to fall in the 5-10 Gyr range. In broad terms, the distributions plotted in the upper two panels of Figure 10 are suggestive of a roughly constant star formation rate over the history of the Galactic disk. This is consistent with the conclusions drawn from a number of other studies, notably the analyses by Soderblom, Duncan \& Johnson (1991) Rocha Pinto et al (2000) and Gizis, Reid \& Hawley's (2003) of the distribution of chromospheric activity in late-type dwarfs.

In sharp contrast to these results, the age distribution of stars known to have planetary companions is stronly slanted toward young and intermediate-age stars (Figure 10, bottom-left panel). The host stars in this sample are drawn from Butler et al.'s (2006) recent catalogue of nearby exoplanets, which includes data for 182 planets in 154 planetary systems. We plot direct age estimates for the 107 systems included in the VF05 dataset, together with estimated ages (based on [Fe/H]) of a further 23 stars. Metallicities for the latter stars are from Santos \etall (2005, 2006), Ecuvillon \etall (2006) and Kotoneva \etall (2006). Most have super-solar metallicities, leading to formal age estimates $\tau <1$ Gyr. These stars are not included in the statistical analysis.

The median age for the 107 VF05 exoplanet hosts is 3.9 Gyrs, or $\sim0.5$ Gyrs younger than the age of the Earth. The overall age distribution is skewed towards young and intermediate ages with $\sim23\%$ of the sample younger than 2.5 Gyrs. This is not consistent with the recent analysis by Saffe, Gomez \& Chavero (2005), who use chromospheric age indicators to derive a relatively flat age distribution for exoplanet and a median ages between 5.2 to 7.4 Gyrs. We have also used a Kolmogorov-Smirnov test to compare the age distribution of the 107 VF05 exoplanet hosts against the 239 VF05 stars in Sample B (Figure 10, bottom right panel). That comparison indicates that there is the probability is less than 5\% that the two samples are drawn from the same parent population - a suggestive, if not conclusive, statistical result. We consider this result, and the implications for the potential age distribution of nearby terrestrial planets, in the following section.

\section {Discussion}

\subsection {The age distribution of exoplanet host stars}

Our analysis of Samples A and B suggests that the ages of local disk stars are broadly consistent with a flat distribution; that is, with a constant star formation rate. The age of the Galactic disk is usually estimated as 8 to 10 Gyrs (Oswalt et al, 1996; Reid, 2005). Under those circumstances, $\sim$25 to 30\% of stars in the Solar Neighbourhood are likely to have ages less than 2.5 Gyrs. Taking the Earth's atmospheric evolution as a template, terrestrial planetary companions of those stars are likely to have anoxic atmospheres, with a significant methane content and correspondingly weak O$_2$, ozone and H$_2$ O signatures.

If planetary systems were distributed randomly among the nearby stars, then the age distribution of these systems would mirror the stellar age distribution. However, the data plotted in Figure 10 suggest that stars known to host planetary systems are skewed towards younger ages than field distribution. As noted is \S3.2, this disagrees with the conclusions drawn by Saffe \etall (2005) from the distribution of Ca II H \& K activity in 112 exoplanet hosts. Chromospheric activity in FGK dwarfs is powered by the $\alpha-\omega$ rotational dynamo (Babcock, 1961), and is known to decline with time, albeit with significant dispersion that reflects the intrinsic spread in properties among even coeval stars: for example, the activity levels measured for Pleiades FGK dwarfs span almost an order of magnitude (Soderblom \etal, 1993). This problem is particularly acute for older, less active stars: the Sun's `chromospheric age' varies from $\sim3$ to $\sim5.5$ Gyrs over the course of the solary cycle; Pace \& Pasquini (2004) note that stars in the intermediate age clusters IC 4651 and NGC 3680 ($\tau \sim1.7$ Gyrs) have chromospheric activity levels comparable to M67 (($\tau \sim 5$ Gyrs); and Giampapa \etall (2006) have shown that M67 stars, themselves, have chromospheric ages ranging from $\sim$0.5 to $\sim7$ Gyrs. 

Activity levels of the exoplanet host stars are measured using the $R'_{HK}$ emission index, and Saffe \etall (2005) apply calibrations by Donahue (1993; D93) and by Rocha-Pinto \& Maciel (1998; RPM98) to estimate ages. The RPM98 age estimator includes a correction for stellar metallicity. Figure 11 compares age estimates derived for the 112 exoplanet hosts from those calibrations, and matches those results against isochrone ages for 92 stars in common with the VF05 sample. There are clearly substantial differences between the three age estimators. Ages derived from the D93 activity index are poorly correlated with the other calibrations; in particular, the D93 ages are, on average, $\sim1.5$ Gyrs older than the VF05 data, with a disperson of $\pm2$ Gyrs. In contrast, the RPM98-based ages are predominantly younger than $\sim4$ Gyrs. The median ages of the D93-, RPM98- and VF05-based distributions are 4.72, 1.82 and 3.7 Gyrs, respectively. 

In their analysis, Saffe \etall favour results derived from the D93 calibration, based largely on the agreement with the distribution derived for the same stars using isochrine-based ages from Nordstr\"om \etall (2004). In general, ages derived from well-calibrated stellar photospheric properties are likely to be more reliable than those based on the characteristics of the thin, magnetically powered chromospheric layers. However, we have already demonstrated that there are problems with the NO4 analysis (see \S3.1, and Figures 1, 7 and 9). Taking due account of the inherent uncertainties, it is likely that the VF05 analysis provides more reliable age estimates for the exoplanet host stars.

The preponderance of exoplanet host stars younger than $\sim5$ Gyrs probably stems from the observed increase in planetary frequency among higher metallicity stars (Gonzalez, 1998; Santos \etal, 2001; Reid, 2002; Fischer \& Valenti, 2005; Ecuvillon \etal, 2005). Most analyses agree that this tendency reflects the fact that planets (even gas giants) are made of `metals'; consequently, their formation is favoured in high metallicity systems. However, some recent theoretical studies (e.g. Ida \& Lin, 2004; Benz et al, 2006) and observations (Udry et al, 2006) suggest that this trend weakens significantly for lower-mass giant planets. How are we to extrapolate from these results to estimate the potential frequency of terrestrial planets? Answering that question depends crucially on the assumptions made regarding the chemical composition of the recently discovered low-mass exoplanets.

There are four distinct types of planetary body within the Solar System: the Jovian gas giants, Jupiter and Saturn, which are $\sim90\%$ hydrogren and helium, $\sim5-10\%$ volatiles and $<1\%$ (i.e. $<3 M_{Earth}$) refractory materials (Owen \& Encranz, 2003); the ice giants, Uranus and Neptune, which ar volatile-rich, with $\sim15\%$ H \& He, 60-70\% volatiles and 15-25\% (2-4 $M_{Earth}$) refractory materials (Guillot, 2005); the terrestrial planets, which are composed predominantly of refractory materials; and the ice dwarfs, like Pluto, Charon and the other Kuiper Belt objects. Recent exoplanet discoveries have extended detections to companions with masses 15-30 M$_{Earth}$, similar to Neptune and Uranus. However, all of these new discoveries lie within ~0.25 AU of the parent star, and are therefore unlikely to be ice giants, unless those planetary systems have experienced radical migration. If the Neptune-mass exoplanets are (compositionally) Jovian analogues, the total mass of refractory elements is $<0.2 M_{Earth}$. The low metal content would account for the weaker correlation with the metallicity of the host stars. 

Resolving this question is beyond the scope of this paper, and, indeed, probably beyond the reach of current observations, at least until suitable transiting exoplanets are uncovered. For present purposes, we can set broad limits on the potential frequency of terrestrial exoplanets. If the formation is independent of the metallicity of the host star, then the age distribution will mirror the underlying stellar age distribution, and 25-30\% of local planetary systems are likely to have ages younger than $\sim2.5$ Gyrs. On the other hand, if terrestrial planet formation is favoured in metal-rich protostellar systems, leading to an age distribution similar to that of known exoplanet systems (Figure 10), then half of the nearby Earth analogues could be younger than 2.5-3 Gyrs.

\subsection{The age distribution of hot jupiters }

The VF05 dataset provides isochrone-based age estimates for 107 exoplanet host stars. This relatively large sample allows us to examine the age distributions of well-chosen subsets. To that end, Figure 11 plots the measured masses of planetary companions, $M_2 \sin{i}$ (where $i$ is orbital inclination), and projected orbital semi-major axis, $a \sin{i}$, as a function of [Fe/H] and age. We distinguish between the closest planet and other companions in multi-planet systems. The uppermost panel shows the companion mass distribution as a function of semi-major axis: the majority of sub-jovian mass planets currently known lie at smaller separations, as one would expect for a radial velocity-selected sample.

The exoplanets plotted in Figure 12 show a smooth distribution in the mass/[Fe/H] and mass/age planes. We note that the lowest mass planets are companions to some of the oldest stars in the sample. In contrast, the semi-major axis distribution appears to be bimodal, with a broad minimum centred at $\log (a \sin{i}) \sim -0.4$, or $\sim0.4$ AU. Udry, Mayor \& Santos (2003) have commented on this bimodality in the log(period) plane. The increased numbers of detected exoplanets at $\log (a \sin{i}) < -1$ likely reflects both the higher sensitivity of radial velocity programs to low-mass companions at small separations, and the recent initiation of surveys such as the N2K (Fisher \etal, 2006) that are directed specifically towards finding short-period systems.

We have divided the VF05 exoplanet sample into stars known to have a planetary companion within 0.4 AU (36 stars, $\langle 0.126 \rangle$ AU; the {it near} sub-sample), and stars that lack such companions (71 stars, $\langle 2.00 \rangle$ AU; the {\it far} sub-sample). Figure 13 plots both differential and cumulative age distributions of the two sub-samples. The age distribution of the {\it near} sub-sample is significantly flatter than the {\it far} sub-sample, lacking the broad peak between 2 to 5 Gyrs. A Kolmogorov-Smirnov test indicates that the there is less than 5\% probability that the two distributions are drawn from the same parent population. 

What is the origin of this difference in the age distributions? One possible explanation is an $M_2 \sin{i}$-metallicity correlation: given that metal-poor stars are, on average, older than metal-rich stars (from the AMR plotted in Figure 7); and that the {\it near} sub-sample of exoplanet hosts extends to lower mass companions; then, if lower metallicity stars are only capable of forming lower-mass planets, the {\it near} sub-sample is likely to include older stars than the {\it far} sample. However, Figure 13 includes a plot of the age-[Fe/H] distribution of exoplanet host stars, where we distinguish between stars in the {\it near} and {\it far} sub-samples; if anything, the {\it near} sub-sample includes fewer metal-poor stars than the {\it far} sub-sample. Indeed, the majority of planetary systems (and almost all stars in the {\it near} sub-sample) have above-average metallicities at all epochs. This suggests that the flatter age distribution of hot jupiters is not an observational selection effect.

Dynamical interactions between planets in multi-planet systems are expected to lead to secular evolution of the orbital parameters. Indeed, particular attention has focused on the potential consequences of these effects for hot jupiters (e.g. Adams \& Laughlin, 2006a, b). However, the flat age distribution observed for the {\it near} sub-sample strongly suggests little depletion over a period of $\sim10$Gyrs, implying that a significant number of these systems are stable on those timescales. On the other hand, the {\it far} sub-sample shows a distinct turnover in numbers at $\tau >5$ Gyrs, perhaps indicating that secular orbital evolution on Gyr-timescales {\sl is} important for these systems (see, for example, Gomes \etal, 2005). These issues can be addressed in more detail both through the identification of new exoplanet systems, and the calculation of reliable, self-consistent age estimates for host stars currently lacking such data.

\subsection {Planet formation in the thick disk}

Although most stars in the Solar Neighbourhood are members of the Galactic disk population, a significant minority (between 5 and 10\%) is drawn from the thick disk. Originally identified by Gilmore \& Reid (1983) as a density excess at moderate heights (1-3 kpc) above the Plane, more recent spectroscopic analyses (e.g. Fuhrmann, 1998, 2004; Prochaska, 2000) have shown that thick disk stars, like Galactic halo stars, have enhanced abundances of $\alpha$ elements (Mg, Ti, O, Ca, Si). These abundance ratios are characteristic of the nucleosynthetic products of Type II supernovae. This implies that the thick disk, like the halo, formed over a relatively short timescale (less than 1-2 Gyrs), before Type I supernovae could increase the iron abundance and decrease the [$\alpha$/Fe] ratio (Matteucci \& Greggio, 1986). 

With abundances in the range $-1 < [Fe/H] < -0.2$ (Figure 14, lower panel), the thick disk clearly formed after the halo, but before the bulk of the Galactic disk. Recent investigations (e.g. Bensby, 2004; Reid, 2005) tend to favour its origin as a consequence of dynamical excitation of a pre-existing thin disk by a major merger early in the Milky Way's history ($t_L > 8$ Gyrs).

Did thick disk stars form planetary systems? As originally discussed by Reid (2006), the completion of several chemical abundance analyses based on high-resolution spectra provides an opportunity to address this question. Besides the Valenti \& Fischer (2005) dataset, which includes abundance measurements for titanium, an $\alpha$ element, Gilli \etall (2006) measure Ca, Mg and Ti abundances for 101 exoplanet hosts; Santos \etall (2006) measure detailed abundances, including Mg, O and Ti, for 6 transiting planets; and Ecuvillon \etall (2006) present O abundances for 96 host stars. Figure 14 combines these datasets, including all 107 VF05 stars, 6 stars from Gilli \etal, 2 stars from Santos \etall and 3 stars from Ecuvillon \etal.  We use Furhmann's (1998) data for nearby stars as a population template. 

Five exoplanet host stars in the present sample meet Fuhrmann's thick disk criterion, with [$\alpha$/Fe]$>0.2$ (Table 2). All five are also relatively metal-poor (for disk dwarfs), $[Fe/H] < -0.3$, and have relatively high space motions with respect to the Sun. The most metal-poor star is HD 114762, which also has a high-mass companion, $M_2 sin(i) = 11.7 M_J$. The system may be observed close to pole-on (Latham \etal, 1989), in which case the companion is likely to be a brown dwarf. HD 111232 b is also relatively high mass, $M_J=6.24 M_J$, but the remaining three stars have companions with $M_2 sin(i) < 1 M_J$, and therefore almost certainly have planetary systems. This clearly demonstrates that, even though most nearby planetary systems are younger than the Sun, planet formation was underway within $\sim1$ Gyr of the formation of the Milky Way.

\subsection { Stellar ages and target selection for the Terrestrial Planet Finder }

How might these results affect the observational strategies adopted by programs that search for terrestrial companions of nearby stars? As a specific example, we can consider the 136 stars identified as prime targets for the Terrestrial Planet Finder (Brown, 2005, http://sco.stsci.edu/starvault/, hereinafter the B136 sample). Those stars are chosen, and ranked in priority, based primarily on the probability of detecting terrestrial planets in the conventional Habitable Zone. The selection criteria include distance ($d<30$ pc), evolutionary state (location on the main sequence), lifetime ((B-V)$>$0.3) and the absence of stellar companions within a 10-arcsecond radius. 

Eighty-seven of the B136 sample are included in the VF05 dataset (most of the non-VF05 stars are F-type dwarfs with M$_V < 3.5$). Figure 15 plots the age-metallicity distribution for these stars, together with colour-magnitude and M$_V$-metallicity diagrams for all 136 stars. Nearly one-third of the VF05 subset (24 of 87 stars) is younger than 2.5 Gyrs, and, taking Earth as a template, any terrestrial companions might be expected to lack O-rich atmospheres. Moreover, almost three quarters are younger than 4 Gyrs; Earth analogues in those systems may well lack significant surface vegetation and the corresponding spectroscopic `red edge' biosignature. Indeed, insofar as the Solar System can be taken as a universal template, Martian analogues in younger systems might still retain water-rich atmospheres, so searches for biosignatures matching present-day Earth should pay careful attention to faint sources near the outer edge of the habitable zone.

Searches for Earth analogues in nearby star systems must take due account of the potential evolution of atmospheric chemistry, and target an appropriate range of biosignatures in searching for habitable exoplanets. Some strategies take this issue into account by searching explicitly for older (oxygen-rich) terrestrial planets; for example, the HabCat systems identified by Turnbull \& Tarter (2003a, b) are limited to main-sequence stars likely to be older than 3 Gyrs.

Measuring accurate ages for isolated stars is difficult, and there are a variety of age indicators with different degrees of reliability (Turnbull \& Tarter, 2003a). In general, the uncertainty in age increases with increasing age; thus, the impact of these uncertainties is muted if the prime concern is placing stars in broad categories reflecting Earth's evolution (for example, ages $<0.5$, $0.5 - 2$, $2-5$ and $>5$ Gyrs). Of the various methods employed to estimate stellar ages, isochrone matching, using parameters derived from analyses of high-resolution spectra, is likely to be more reliable than either photometric analyses or investigations based on rotation or (intrinsically variable) chromospheric activity. Extending the VF05 dataset to include all potential TPF targets would provide a self-consistent set of age estimates, and should be a high priority for preparatory TPF science.

\section {Summary and conclusions}

We have utilised the results from Valenti \& Fischer's (2005) detailed analysis of high-resolution spectroscopy of 1039 nearby solar-type stars to probe the intrinsic properties of local disk stars. We have identified appropriate subsets of long-lived FGK dwarfs from this parent sample. Using the Hipparcos catalogue as a reference, we have defined volume-complete samples of main-sequence stars with $4 < M_V < 6$ and $d<30$ parsecs, and $4 < M_V < 6.5$ and $d < 25$ parsecs. Sixty percent of the stars in those samples are included in the VF05 dataset; we have compiled metallicity information (but not ages) from the literature for the remaining stars.  

Our main results are as follows:
\begin{itemize}
\item The Valenti \& Fisher dataset, which represents the main target list for the Lick/Keck/AAT Planet Search programs, includes a higher proportion of metal-rich stars than an unbiased sample of the local disk population.
\item The Valenti \& Fisher data reveal a clear age-metallicity relation for local Galactic disk stars. The trend with age can be represented as a linear relation (with substantial dispersion), with the mean metallicity increasing from $\sim-0.25$ dex at age $t_L=10$ Gyrs to $\sim+0.15$ dex at the present day. 
\item The overall age distribution of local disk stars is broadly consistent with a uniform star-formation rate over the history of the Galactic disk. However, the age distribution of stars that are currently known to have (gas giant) exoplanet companions is strongly skewed to ages younger than 5 Gyrs, presumably reflecting the higher frequency of those systems among metal-rich stars.
\item We have divided the exoplanet hosts into stars with known planetary companions with $a < 0.4$AU, and stars where the nearest known companion lies at $a > 0.4$AU. The former sample has a flat age distribution, while the latter has a strong peak at ages between 2 and 5 Gyrs. 
\end{itemize}

Age is a crucial parameter in assessing the likely atmospheric composition of terrestrial exoplanets. At the present juncture we lack direct detections of any such planets. If terrestrial planet formation is independent of the metallicity of the host star, then at least 25\% of local systems are expected to be younger than $\sim2.5$ Gyrs, a period during which Earth's atmosphere was anoxic. However, if the terrestrial planetary systems follow an age distribution similar to the known exoplanet host stars, then 40-50\% of the Earth-analogues in the Solar Neighbourhood could be younger than 2.5 Gyrs. A significant fraction of the nearby stars likely to be TPF targets still lack thorough high-resolution spectroscopic analyses. We strongly advocate acquiring the appropriate data and undertaking those analyses as an essential precursor to defining a final TPF target list.

\acknowledgements

The authors thank Dave Soderblom, Eric Ford and Lisa Kaltenegger for useful comments and suggestions. We also acknowledge useful comments and suggestions from the anonymous referee that highlighted some areas requiring clarification in the original manuscript. 

{}

\clearpage

\begin{deluxetable}{crrrrrrrr}
\tabletypesize{\scriptsize}
\tablewidth{0pt}
\tablecolumns{8}
\tablecaption{Comparison between {\sl uvby} and VF05 metallicity calibrations}
\tablehead{ \colhead {} & \colhead {Haywood} & \colhead {} & \colhead{N)4} \\
\colhead{$(b-y)$} & \colhead{$\delta$[Fe/H]} & \colhead{$\sigma_{\delta}$} &
\colhead{$\delta$[Fe/H]} & \colhead{$\sigma_{\delta}$} & 
 \colhead{N}}
\startdata
0.35 {--} 0.4 & 0.015 & 0.13 & 0.09 & 0.11 & 197 \\
0.4 {--} 0.45 & -0.002 & 0.09 & 0.07 & 0.09 & 181 \\
0.45 {--} 0.5 & 0.045 & 0.12 & 0.05 & 0.11 & 77 \\
0.5 {--} 0.55 & -0.041 & 0.15 & 0.04 & 0.15 & 40 \\
\enddata
\tablecomments{ Columns 2 and 3 list the mean difference and the rms dispersion between the VF05 metallicities and those derived using the Haywood (2002) {\sl uvby} metallicity calibration; Columns 4 and 5 list the same parameters in a comparison between the VF05 dataset and the N04 calibration. Column 6 lists the total number of stars in each colour bin. The individual datapoints are plotted in Figure 1.}
\end{deluxetable}

\clearpage

\begin{deluxetable}{crcrrrrrrrr}
\tabletypesize{\scriptsize}
\tablewidth{0pt}
\tablecolumns{10}
\tablecaption{Thick disk exoplanet host stars}
\tablehead{ \colhead {Name} & \colhead {M$_V$} & \colhead {Sp. type} & \colhead{[Fe/H]} & \colhead{U kms$^{1}$} & \colhead{V kms$^{1}$} & \colhead{W kms$^{1}$} & \colhead{$M_2 \sin{i} \ M_J$} & \colhead{$a \sin{i}$ AU} & \colhead{Refs.}}
\startdata
HD 4308 & 4.85 & G5V & -0.31 & 52 & -111 & -29 & 0.0467 & 0.118 & 1, 2 \\
HD 6434 & 4.69 & G2/3V& -0.52& 85 & -67 & -3 & 0.397 & 0.142 & 3, 4\\
HD 37124$^a$ & 5.07 & G4V & -0.44& 22 & -47 & -44 & 0.638 & 0.529 & 1, 5\\ 
&  &  & & &  & & 0.624 & 1.64 & \\
&  &  & & &  & & 0.683 & 3.19 & \\
HD 111232 & 5.29 & G8V & -0.36 & 59 & -84 & 5 & 6.84 & 1.97 & 6, 4\\
HD 114762 & 4.26 & F9V/VI& -0.65 & -82 & -70 & 59 & 11.7 & 0.363 & 1, 7 \\
\enddata
\tablecomments{a: HD 37124 has three known planetary-mass companions.\\
References: \\
1. Valenti \& Fischer, 2005; 2. Udry \etal, 2006; 3. Ecuvillon \etal, 2006; 4. Mayor \etal, 2004; 5. Vogt \etal, 2005; 6. Gilli \etal, 2006; 7. Latham \etal, 1989 }
\end{deluxetable}

\clearpage

\centerline {Figure captions}

\figcaption{ The uppermost panel compares metallicity estimates derived from {\sl uvby} photometry using the prescriptions provided by Nordstr\"om et al (2004, N04) and Haywood (2002, H). The differences, in the sense [Fe/H]$_H$-[Fe/H]$_N$, are plotted as a function of $(b-y)$ colour. The two lower panels compare the {\sl uvby}-based metallicities against the spectroscopic [Fe/H] values derived by Valenti \& Fischer (2005). The open circles plot the mean offset as a function of $(b-y)$; as Table 1 shows, the dispersions, $\sigma_{[Fe/H]}$, about the mean are similar, but there is clearly better agreement between the Haywood calibration and the VF05 dataset.}

\figcaption{ The upper panel shows the (M$_V$, distance) distribution of the 1039 stars in the VF05 sample; the lower panel plots the (M$_V$, (B-V)) colour-magnitude diagram for the stars in the sample.}

\figcaption{ The completeness of the VF05 sample as a function of distance. We divide the sample into 0.5-magnitude bins in M$_V$ and take the Hipparcos catalogue as our reference; each panel plots the fractional contribution of VF05 stars to the Hipparcos sample as a function of distance. It is clear that the overall completeness declines significantly at M$_V > 6.5$.}

\figcaption{ A comparison between the (M$_V$, (B-V)) and (M$_V$, (b-y)) colour-magnitude diagrams for VF05 and non-VF05 stars in Sample A: Hipparcos catalogue stars with $4 < M_V < 6$, $d<30$ pc. As described in the text, there are 297 stars in the left-hand panels (VF05 subset), and 268 stars in the right-hand panels.}

\figcaption { A comparison between the (M$_V$, (B-V)) and (M$_V$, (b-y)) colour-magnitude diagrams for VF05 and non-VF05 stars in Sample B: Hipparcos catalogue stars with $4 < M_V < 6.5$, $d<25$ pc. As described in the text, there are 239 stars in the left-hand panels (VF05 subset; 183 in common with Sample A), and 268 stars (133 in common with Sample A) in the right-hand panels.}

\figcaption { A comparison between the abundance distribution of the VF05 and non-VF05 datasets from samples A and B. This comparison is based primarily on Haywood-calibrated {\sl uvby} metallicity estimates for {\sl both} samples (see text for full details). The left-hand panels plot the differential distributions, where the solid line plots the VF05 sample and the dotted line the non-VF05 sample. The right hand panels plot the metallicity distributions in cumulative form, using the same conventions (solid line for VF05). The solid points show the fractional contribution of the VF05 dataset to the full sample as a function of metallicity (i.e. f=0.5 indicates that half of the stars from Sample A (or B) at that particular metallicity are in the VF05 dataset). The VF05 sub-samples include a higher proportion of the metal-rich stars in both Samples A and B. The lowest panel plots the metallicity distribution of stars known to have planetary-mass companions (the solid histogram shows the contribution from subgiant stars).}

\figcaption{ The age-metallicity relation for the local Galactic disk: The top-left panel plots data for the VF05 stars from Sample A; the lower-left panel shows data for the VF05 stars from Sample B; the top-right panel plots results for the Edvardsson et al (1993) dataset; and the lower-right panel plots the AMR defined by the Nordstr\"om et al (2004) dataset. In each case, the solid pentagon marks the location of the Sun. The two upper panels also show the best-fit linear and second-order relations for the VF05 data; the large crosses in the upper-right panel plot the AMR derived by Rocha-Pinto et al (2000); and the errorbars in the lower-left panel provide an indication of the range of uncertainties associated with the VF05 age estimates.}

\figcaption {Comparison between the VF05, E93 and N04 analyses. The left hand panels compare metallicities and ages for 26 stars in common between the VF05 and E93 samples; there is a small systematic offset in [Fe/H], with VF05 stars $\sim0.1$ dex more metal rich, but the ages are in reasonable agreement. There is larger scatter between the VF05 and N04 metallicities (as illustrated in Figure 1); however, there is no obvious correlation between the ages derived in the two analyses.}

\figcaption {Theoretical models and the age distribution: the left-hand panel superimposes predictions from the Yale-Yonsei models on the observed colour-magnitude distribution of stars in Sample B. The solid points plot the predicted locations of stars with [Fe/H]=+0.38 (Z=0.04) and masses 0.92, 1.01 and 1.20 $M_\odot$; the solid squares plot data for [Fe/H]=0.04 (Z=0.00) and masses 0.84, 0.92, 1.01 and 1.20 $M_\odot$; and the open circles plot data for [Fe/H]=-0.43 (Z=0.007) and masses 0.76, 0.84, 0.92, 1.01 and 1.20 $M_\odot$. In each case, [$\alpha$/Fe]=0.0, and data are plotted for ages 0.4, 1.0 and 5.0 Gyrs. The uppermost right-hand panel plots the (age, M$_V$) distribution for the all 239 VF05 stars in Sample B; the middle panel plots the ([Fe/H], M$_V$) distribution for the same dataset; and the lowest panel plots the age-metallicity relation for VF05 stars in Sample B with M$_V>5.0$ (i.e. stars with lifetimes longer than the age of the disk). The solid line in the last diagram is the linear AMR listed as equation (1), and the solid hexagon marks the Suns' location.}

\figcaption{ The age distribution of local disk stars: The upper two panels show the age distributions for the volume-complete samples considered in the present study: the shaded histogram plots the summed probability distribution for stars in the VF05 dataset, and the dotted histogram is based on the median ages for those stars; the solid histogram includes non-VF05 stars, whose ages are estimated using the linear AMR plotted in Figure 7. In both cases, the vertical bars mark the median ages for the full sample (solid line) and for the VF05 stars alone (dotted line). The lower left panel plots the age distribution of stars known to have planetary companions: the shaded histogram shows data for 107 VF05 stars with isochrone-based ages; a further 23 stars have age estimates that are based on the linear AMR. The vertical bar (dotted) marks the median age for the VF05 host stars. Finally, the lower right panel compares the cumulative age distributions of the 107 VF05 exoplanet hosts and the 239 VF05 stars from Sample B; a Kolmogorov-Smirnov test indicates that the probability is less than 5\% that the two samples are drawn from the same parent population.} 

\figcaption{ Age estimates for exoplanet host stars: the right-hand upper panels plot  age distributions derived for the 112 stars from Saffe \etall (2005) using the $R'_{HK}$-based age calibrations derived by Donahue (1993) and Rocha-Pinto \& Maciel (1998); the lowest panel shows the age distribution for the 92 stars that are also included in the VF05 dataset. The vertical bars mark the median age for each sample. The left-hand panels show a star-by-star comparison of ages derived using the three techniques.}

\figcaption{ Data for the 107 exoplanet host stars included in the VF05 survey: we show the (projected) companion mass and semi-major axis as a function of both [Fe/H] and age; solid squares plot data for the closest planet in each system, crosses plot data for other companions in multi-planet systems. The uppermost panel shows the companion mass/semi-major axis distribution, where the larger span in mass at small separations reflects the greater sensitivity of radial velocity surveys to close companions. }

\figcaption{ Age distributions for exoplanet host stars. The left-hand panels plot differential distributions, with the uppermost plotting ; as in Figure 8, the hatched histogram plots data for the 107 stars in the VF05 dataset. The middle histogram plots the age distribution for the 38 VF05 stars with planetary companions with $a \sin{i} < 0.4$ AU, the {\it near} sub-sample; the age distribution for the 69 stars in {\it far} sub-sample is plotted in the lower left diagram. The upper right-hand panel shows the cumulative age distributions for the {\it near} (dotted line) and {\it far} (solid line) sub-samples. Finally, the lower-right panel plots the age-metallicity distribution for the 107 exoplanet hosts included in the VF05 sample: stars from the {\it near} sample are plotted as solid squares, and stars from the {\it far} sample as open circles. The solid line marks the linear AMR derived from the volume-limited VF05 sub-sample plotted in Figure 7. }

\figcaption{ Thick disk planetary systems: The lower panel plots $\alpha$-element abundances (as exemplified by [Mg/Fe]) for nearby stars from Fuhrmann (1998), where the open squares are disk dwarfs, the solid squares mark stars identified as members of the thick disk, and four-point stars mark transition objects. The upper panel plots [Ti/Fe]/[Fe/H] data from Valenti \& Fischer's (2005) analysis of stars in the Berkeley/Carnegie planet survey (crosses), together with results from other high-resolution abundance analyses of exoplanet hosts. The solid points mark VF05 stars known to have planetary companions; the open circles plot data from Gilli \etall (2006), Santos \etall (2006) (both [Ti/Fe] abundances) and Ecuvillon \etall (2006) ([O/Fe abundances). The five exoplanet hosts with [$\alpha$/Fe]$>0.2$ are listed in Table 2 and discussed in the text.}

\figcaption{ Colour-magnitude, M$_V$-metallicity and age-metallicity distributions for the 136 stars in Brown's high priority TPF target list. The 87 stars that are included in the VF05 dataset are plotted as solid squares. }

\clearpage

\begin{figure}
\figurenum{1}
\plotone{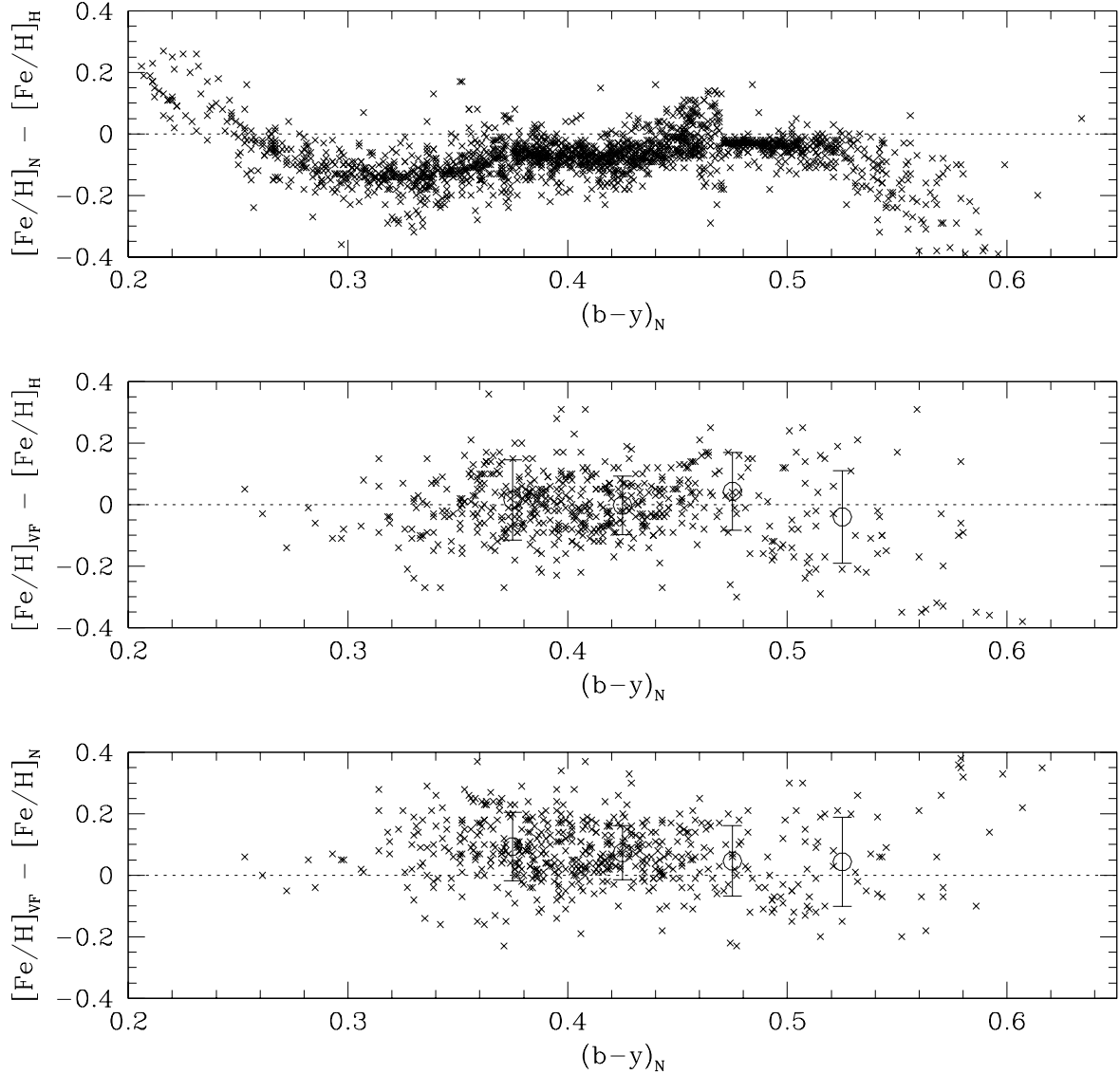}
\caption {The uppermost panel compares metallicity estimates derived from {sl uvby} photometry using the prescriptions provided by Nordstr\"om et al (2004) and Haywood (2002). The differences, in the sense [Fe/H]$_H$-[Fe/H$_N$, are plotted as a function of $(b-y)$ colour. The two lower panels compare the {\sl uvby}-based chemical abundances against the spectroscopic [Fe/H] values derived by Valenti \& Fischer (2005). The open circles plot the mean offset a a function of $(b-y)$; as Table 1 shows, the dispersion, $\sigma_{[Fe/H]}$, about the mean is similar in both cases, but there is slightly better agreement between the Haywood calibration and the VF05 dataset.}
\end{figure}

\begin{figure}
\figurenum{2}
\plotone{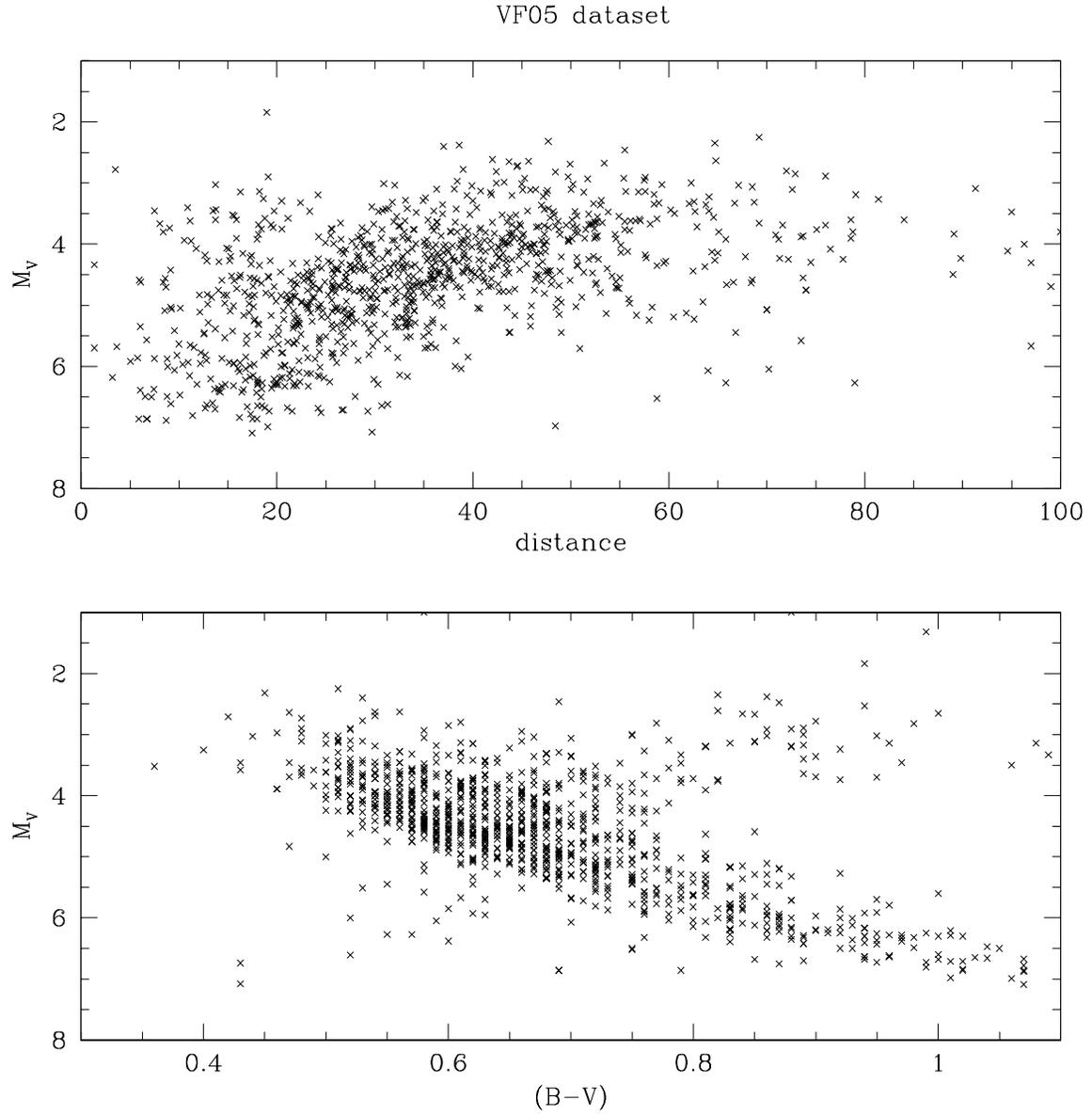}
\caption {The upper panel shows the (M$_V$, distance) distribution of the 1039 stars in the VF05 sample; the lower panel plots the (M$_V$, (B-V)) colour-magnitude diagram for the stars in the sample.}
\end{figure}

\begin{figure}
\figurenum{3}
\plotone{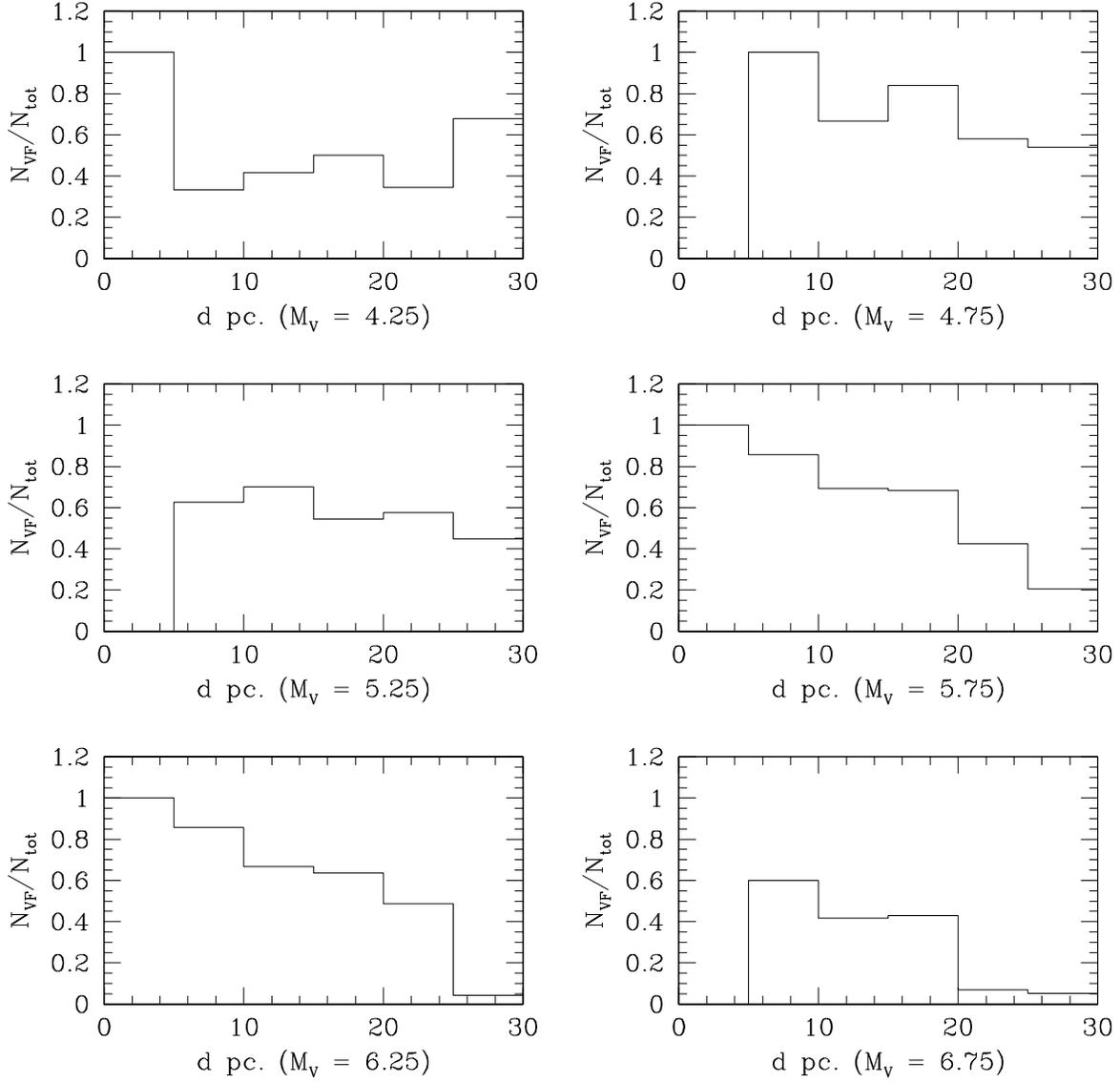}
\caption { The completeness of the VF05 sample as a function of distance; the six panels are divided by absolute magnitude, M$_V$, and show the fraction of stars from the Hipparcos distance-limited sample that are also included in the VF05 dataset. It is clear that completeness declines significant for $d>25$ pc and M$_V > 6$.}
\end{figure}

\begin{figure}
\figurenum{4}
\plotone{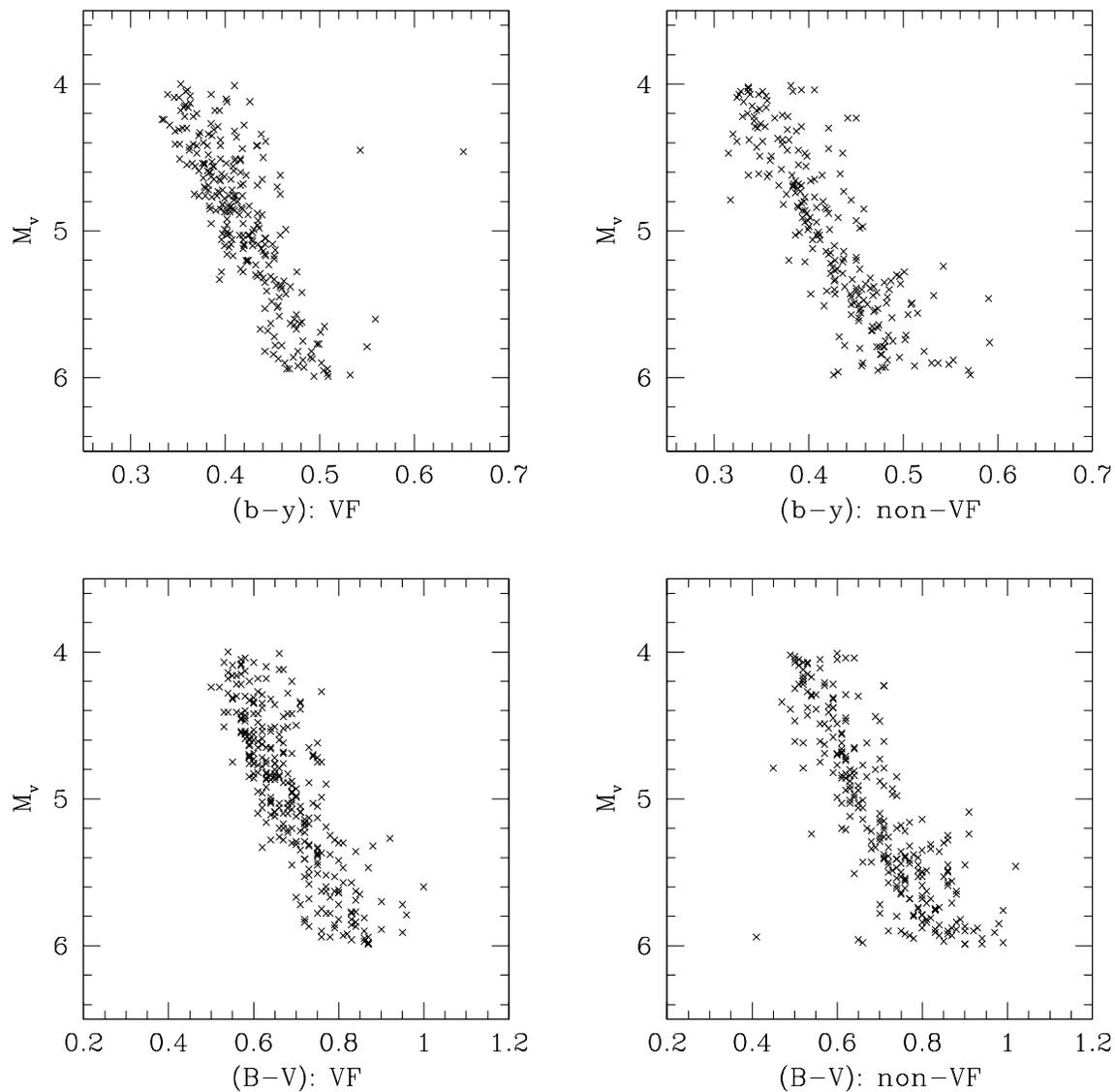}
\caption { A comparison between the (M$_V$, (B-V)) and (M$_V$, (b-y)) colour-magnitude diagrams for stars in Sample A: $4 < M_V < 6$, $d<30$ pc. As described in the text, there are 297 stars in the left-hand panels (VF05 subset), and 268 stars in the right-hand panels.}
\end{figure}

\begin{figure}
\figurenum{5}
\plotone{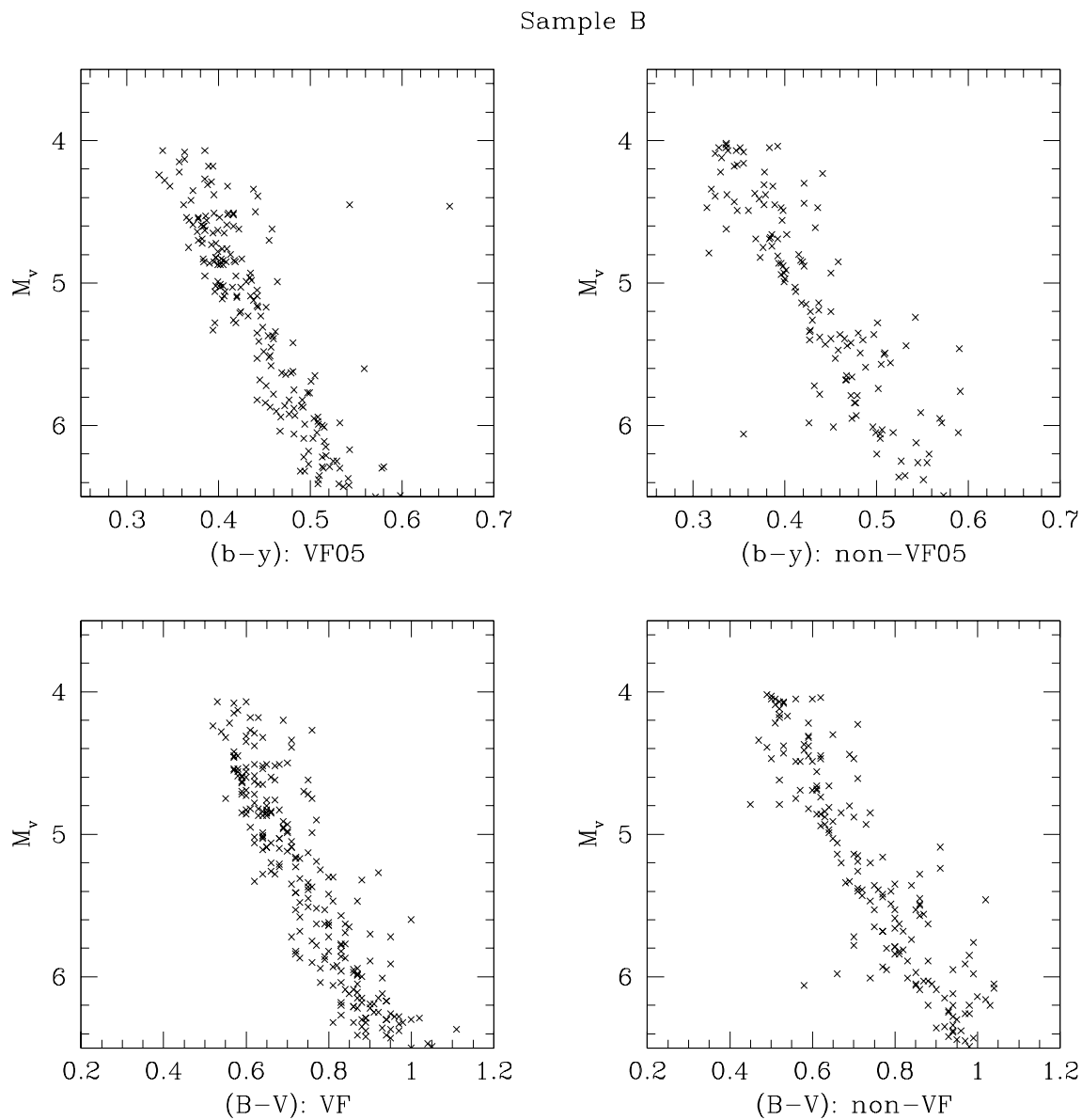}
\caption { A comparison between the (M$_V$, (B-V)) and (M$_V$, (b-y)) colour-magnitude diagrams for stars in Sample B: $4 < M_V < 6.5$, $d<25$ pc. As described in the text, there are 239 stars in the left-hand panels (VF05 subset; 183 in common with Sample A), and 268 stars (133 in common with Sample A) in the right-hand panels.}
\end{figure}

\begin{figure}
\figurenum{6}
\plotone{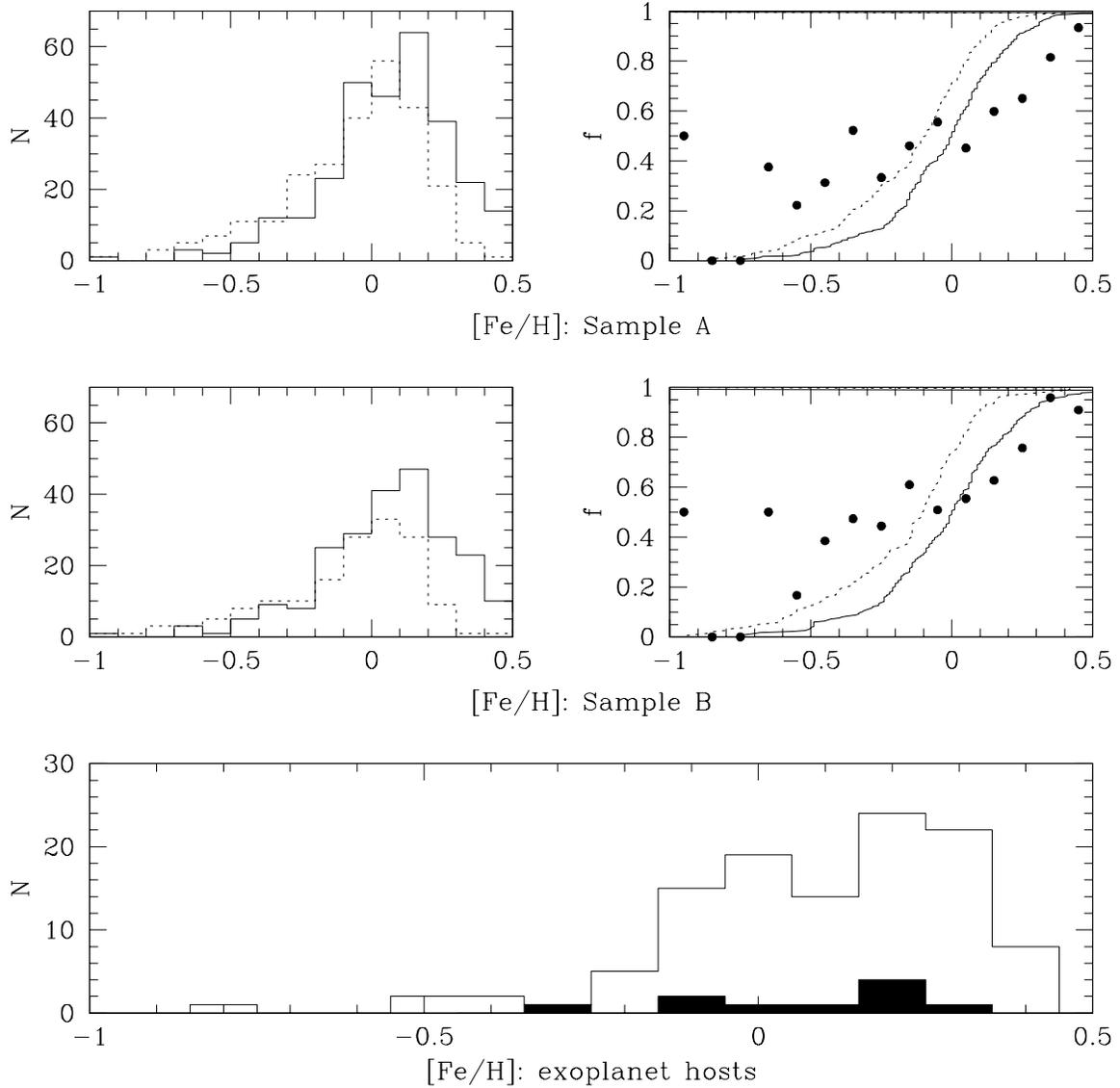}
\caption{ A comparison between the abundance distribution of the VF05 and non-VF05 datasets from samples A and B. This comparison is based primarily on Haywood-calibrated {\sl uvby} metallicity estimates for {\sl both} samples (see text for full details). The left-hand panels plot the differential distributions, where the solid line plots the VF05 sample and the dotted line the non-VF05 sample. The right hand panels plot the metallicity distributions in cumulative form, using the same conventions (solid line for VF05). The solid points show the fractional contribution of the VF05 dataset to the full sample as a function of metallicity (i.e. f=0.5 indicates that half of the stars from Sample A (or B) at that particular metallicity are in the VF05 dataset). The VF05 sub-samples include a higher proportion of the metal-rich stars in both Samples A and B. The lowest panel plots the metallicity distribution of stars known to have planetary-mass companions (the solid histogram shows the contribution from subgiant stars).}
\end{figure}

\begin{figure}
\figurenum{7}
\plotone{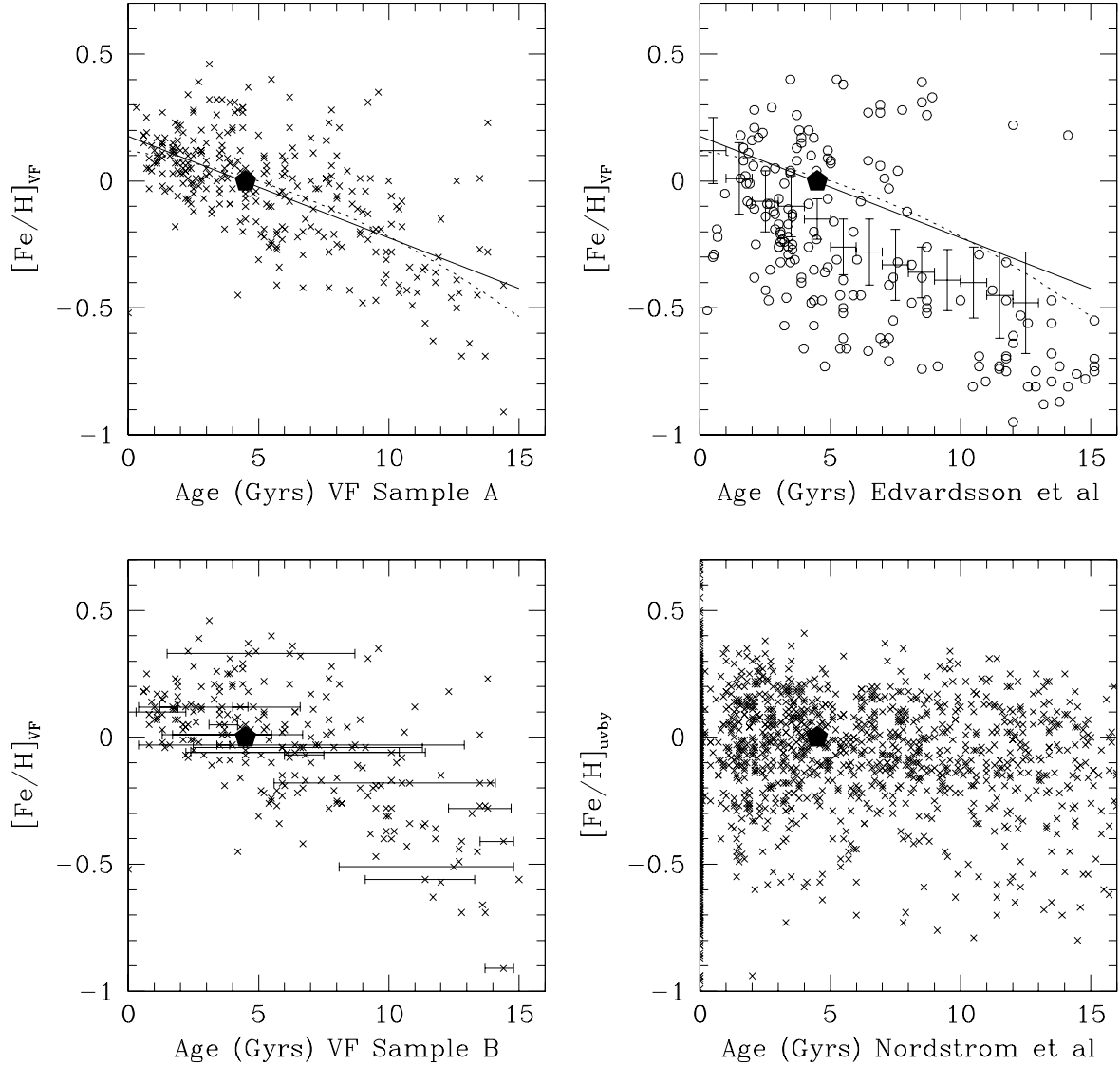}
\caption{ The age-metallicity relation for the local Galactic disk: The top-left panel plots data for the VF05 stars from Sample A; the lower-left panel shows data for the VF05 stars from Sample B; the top-right panel plots results for the Edvardsson et al (1993) dataset; and the lower-right panel plots the AMR defined by the Nordstr\"om et al (2004) dataset. In each case, the solid pentagon marks the location of the Sun. The two upper panels also show the best-fit linear and second-order relations for the VF05 data; the large crosses in the upper-right panel plot the AMR derived by Rocha-Pinto et al (2000); and the errorbars in the lower-left panel provide an indication of the range of uncertainties associated with the VF05 age estimates.}
\end{figure}

\begin{figure}
\figurenum{8}
\plotone{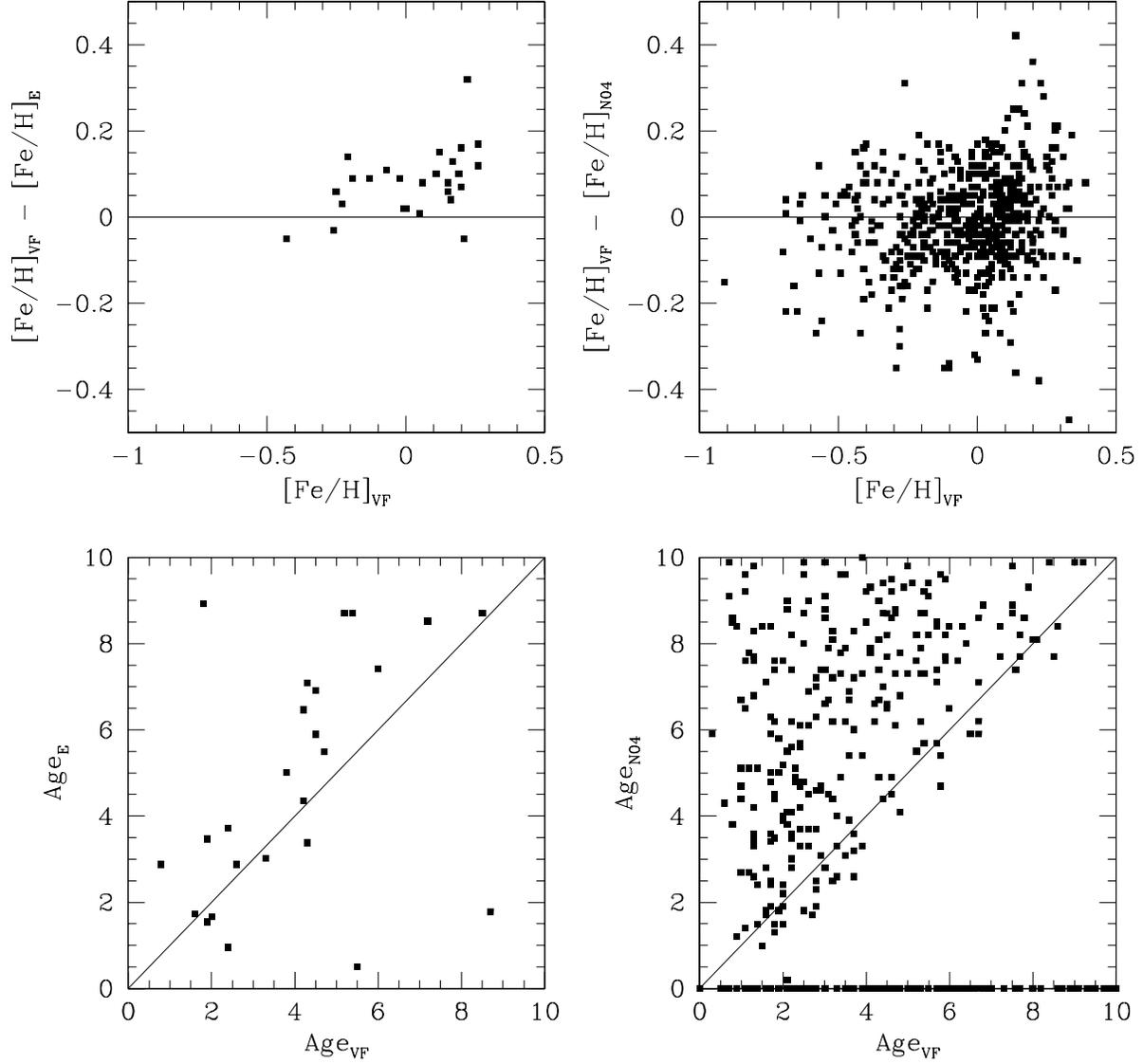}
\caption{Comparison between the VF05, E93 and N04 analyses. The left hand panels compare metallicities and ages for 26 stars in common between the VF05 and E93 samples; there is a small systematic offset in [Fe/H], with VF05 stars $\sim0.1$ dex more metal rich, but the ages are in reasonable agreement. There is larger scatter between the VF05 and N04 metallicities (as illustrated in Figure 1); however, there is no obvious correlation between the ages derived in the two analyses.}
\end{figure}

\begin{figure}
\figurenum{9}
\plotone{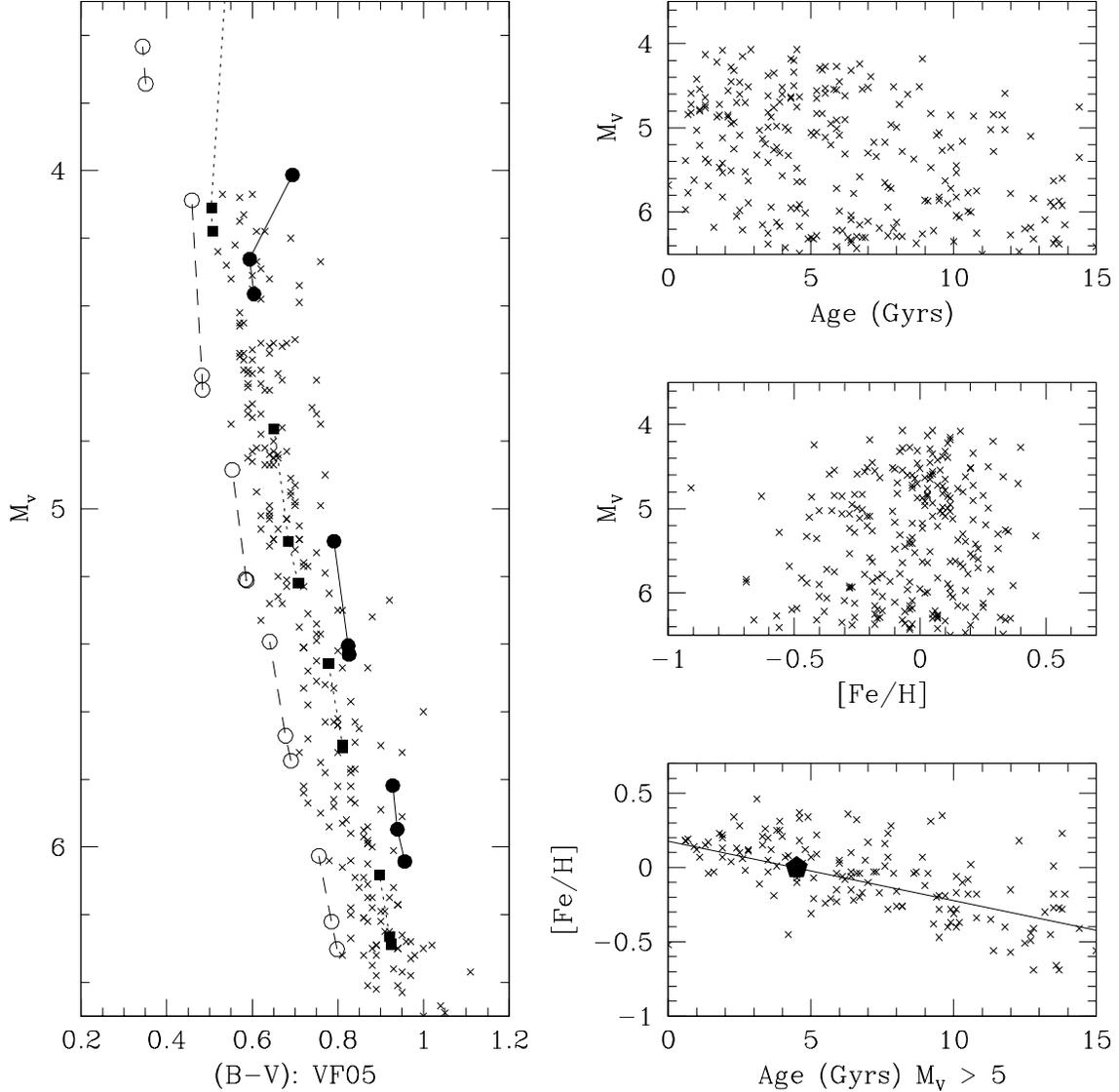}
\caption{Theoretical models and the age distribution: the left-hand panel superimposes predictions from the Yale-Yonsei models on the observed colour-magnitude distribution of stars in Sample B. The solid points plot the predicted locations of stars with [Fe/H]=+0.38 (Z=0.04) and masses 0.92, 1.01 and 1.20 $M_\odot$; the solid squares plot data for [Fe/H]=0.04 (Z=0.00) and masses 0.84, 0.92, 1.01 and 1.20 $M_\odot$; and the open circles plot data for [Fe/H]=-0.43 (Z=0.007) and masses 0.76, 0.84, 0.92, 1.01 and 1.20 $M_\odot$. In each case, [$\alpha$/Fe]=0.0, and data are plotted for ages 0.4, 1.0 and 5.0 Gyrs. The uppermost right-hand panel plots the (age, M$_V$) distribution for the all 239 VF05 stars in Sample B; the middle panel plots the ([Fe/H], M$_V$) distribution for the same dataset; and the lowest panel plots the age-metallicity relation for VF05 stars in Sample B with M$_V>5.0$ (i.e. stars with lifetimes longer than the age of the disk). The solid line in the last diagram is the linear AMR listed as equation (1), and the solid hexagon marks the Suns' location.}
\end{figure}

\begin{figure}
\figurenum{10}
\plotone{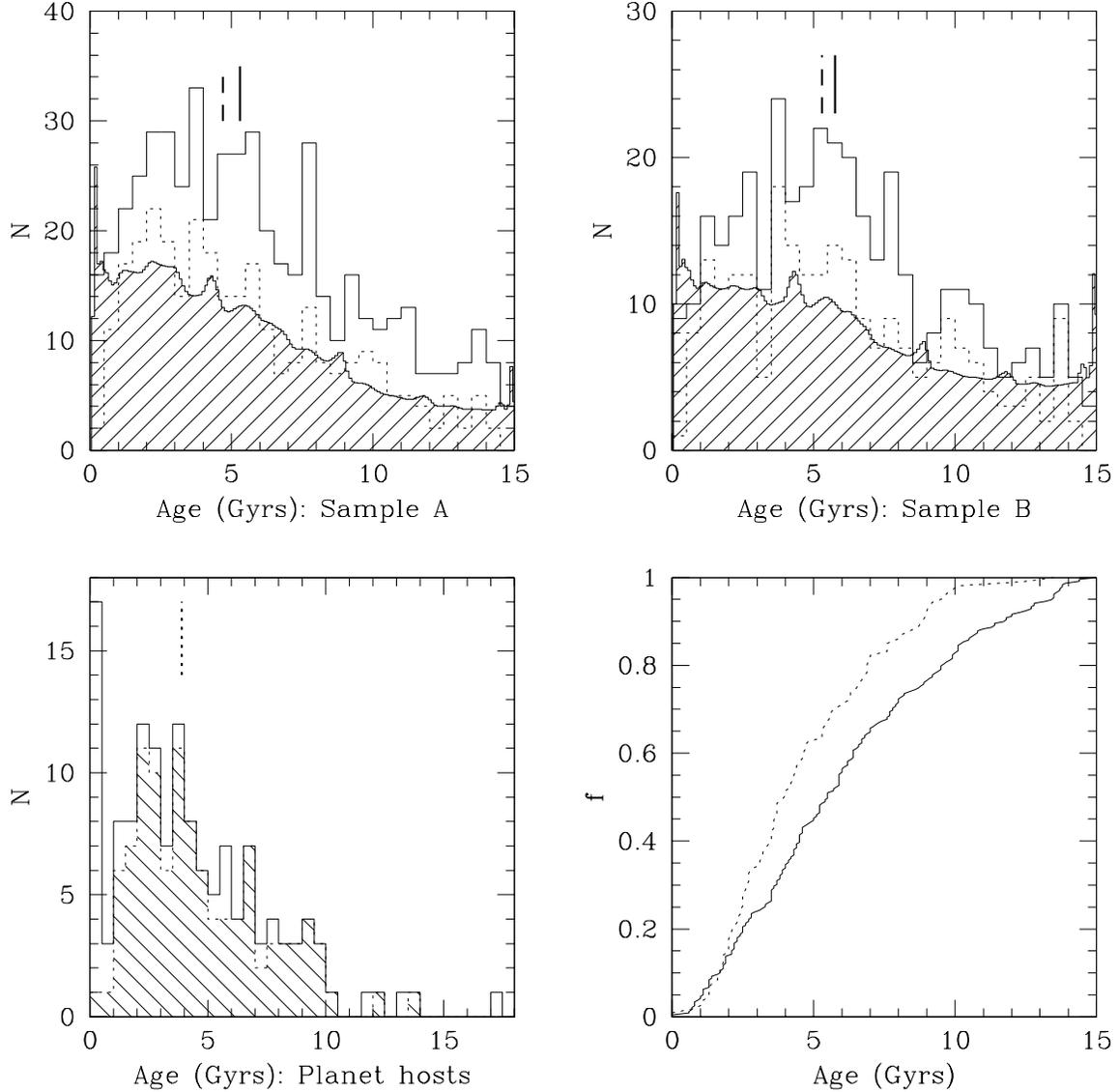}
\caption{ The age distribution of local disk stars: The upper two panels show the age distributions for the volume-complete samples considered in the present study: the shaded histogram plots the summed probability distribution for stars in the VF05 dataset, and the dotted histogram is based on the median ages for those stars; the solid histogram includes non-VF05 stars, whose ages are estimated using the linear AMR plotted in Figure 7. In both cases, the vertical bars mark the median ages for the full sample (solid line) and for the VF05 stars alone (dotted line). The lower left panel plots the age distribution of stars known to have planetary companions: the shaded histogram shows data for 107 VF05 stars with isochrone-based ages; a further 23 stars have age estimates that are based on the linear AMR. The vertical bar (dotted) marks the median age for the VF05 host stars. Finally, the lower right panel compares the cumulative age distributions of the 107 VF05 exoplanet hosts and the 239 VF05 stars from Sample B; a Kolmogorov-Smirnov test indicates that the probability is less than 5\% that the two samples are drawn from the same parent population.} 
\end{figure}

\begin{figure}
\figurenum{11}
\plotone{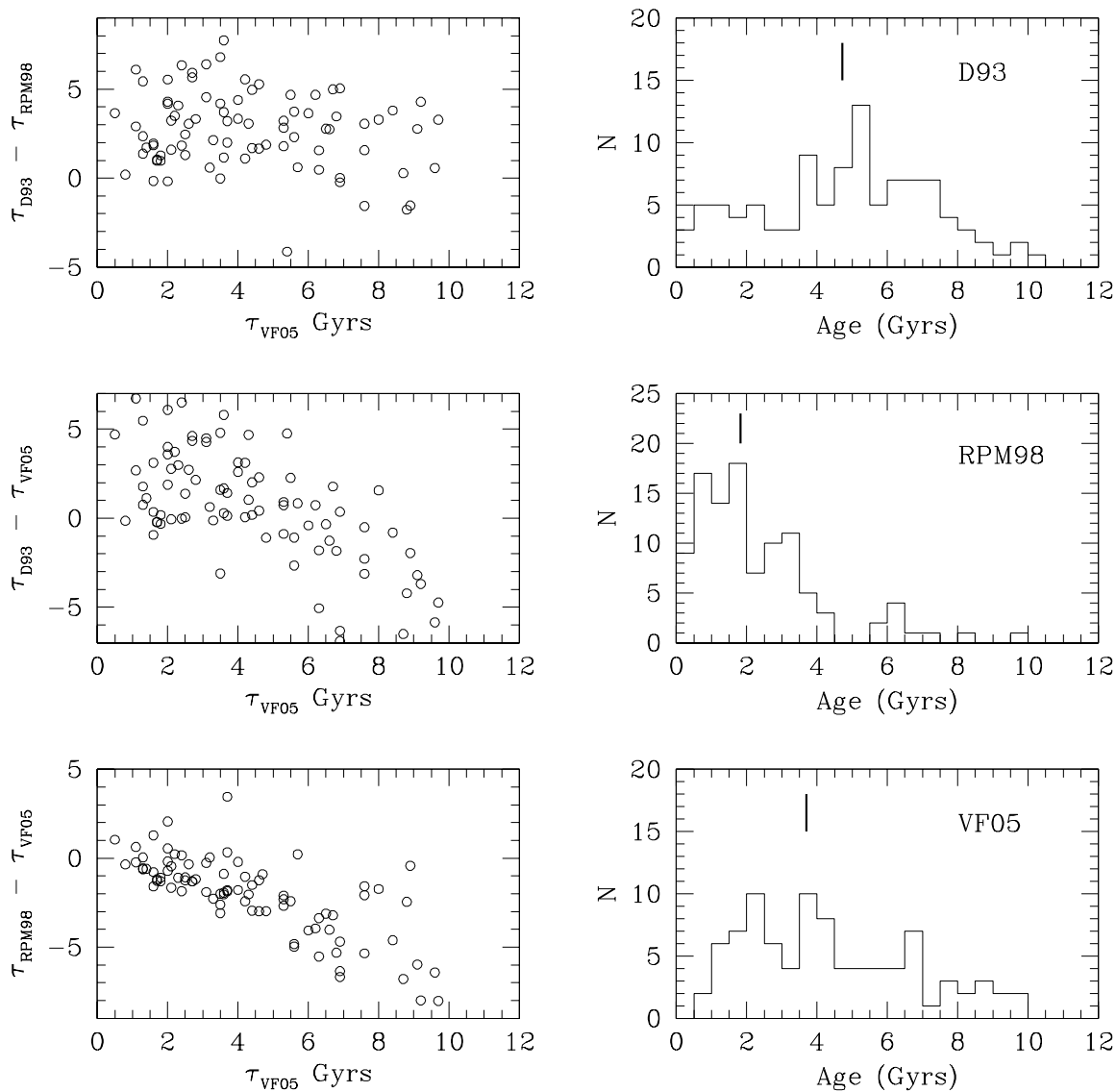}
\caption{ Age estimates for exoplanet host stars: the right-hand upper panels plot  age distributions derived for the 112 stars from Saffe \etall (2005) using the $R'_{HK}$-based age calibrations derived by Donahue (1993) and Rocha-Pinto \& Maciel (1998); the lowest panel shows the age distribution for the 92 stars that are also included in the VF05 dataset. The vertical bars mark the median age for each sample. The left-hand panels show a star-by-star comparison of ages derived using the three techniques.}
\end{figure}

\begin{figure}
\figurenum{12}
\plotone{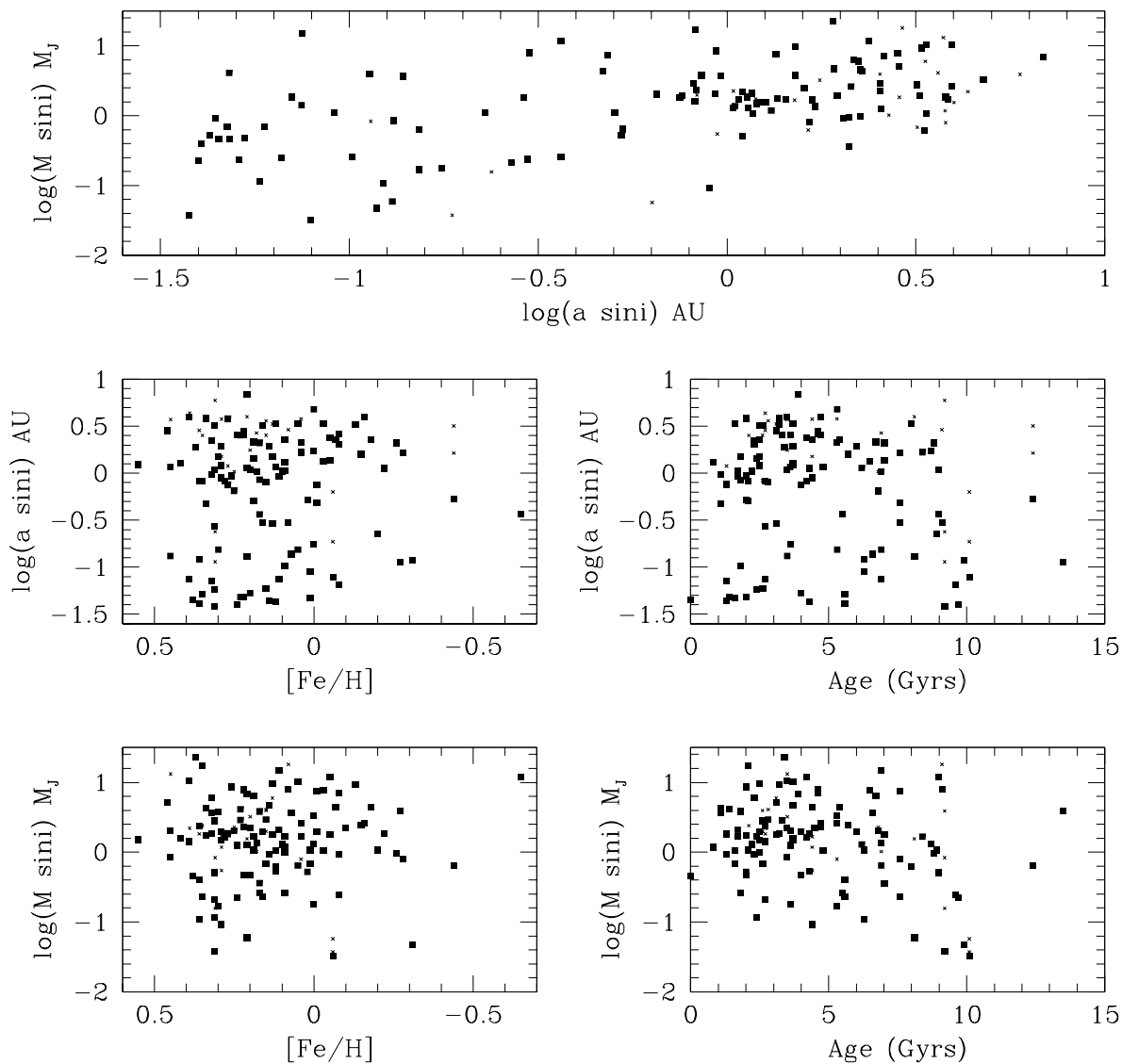}
\caption{ Data for the 107 exoplanet host stars included in the VF05 survey: we show the (projected) companion mass and semi-major axis as a function of both [Fe/H] and age; solid squares plot data for the closest planet in each system, crosses plot data for other companions in multi-planet systems. The uppermost panel shows the companion mass/semi-major axis distribution, where the larger span in mass at small separations reflects the greater sensitivity of radial velocity surveys to close companions. }
\end{figure}

\begin{figure}
\figurenum{13}
\plotone{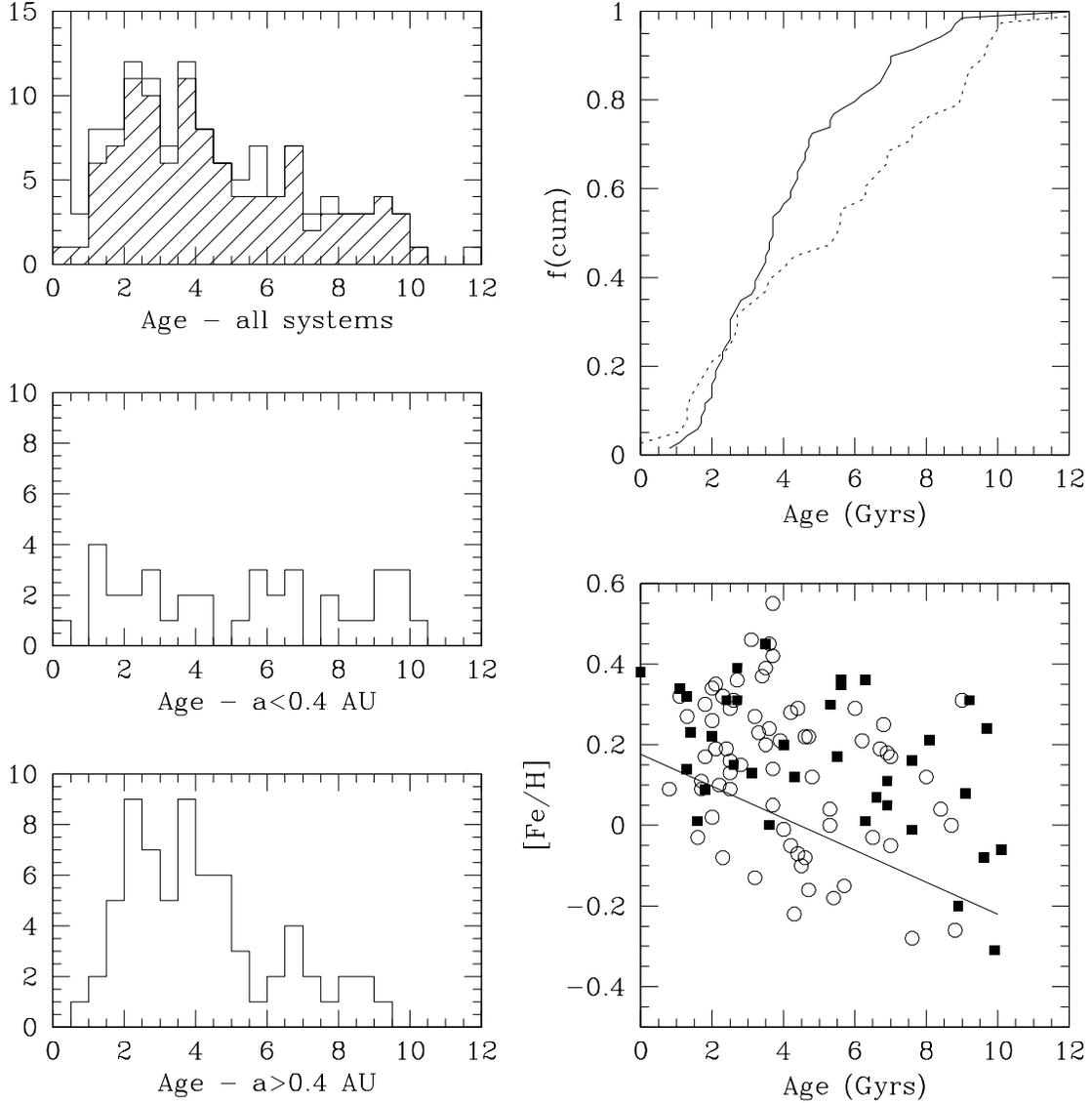}
\caption{ Age distributions for exoplanet host stars. The left-hand panels plot differential distributions, with the uppermost plotting ; as in Figure 8, the hatched histogram plots data for the 107 stars in the VF05 dataset. The middle histogram plots the age distribution for the 38 VF05 stars with planetary companions with $a \sin{i} < 0.4$ AU, the {\it near} sub-sample; the age distribution for the 69 stars in {\it far} sub-sample is plotted in the lower left diagram. The upper right-hand panel shows the cumulative age distributions for the {\it near} (dotted line) and {\it far} (solid line) sub-samples. Finally, the lower-right panel plots the age-metallicity distribution for the 107 exoplanet hosts included in the VF05 sample: stars from the {\it near} sample are plotted as solid squares, and stars from the {\it far} sample as open circles. The solid line marks the linear AMR derived from the volume-limited VF05 sub-sample plotted in Figure 7. }
\end{figure}

\begin{figure}
\figurenum{14}
\plotone{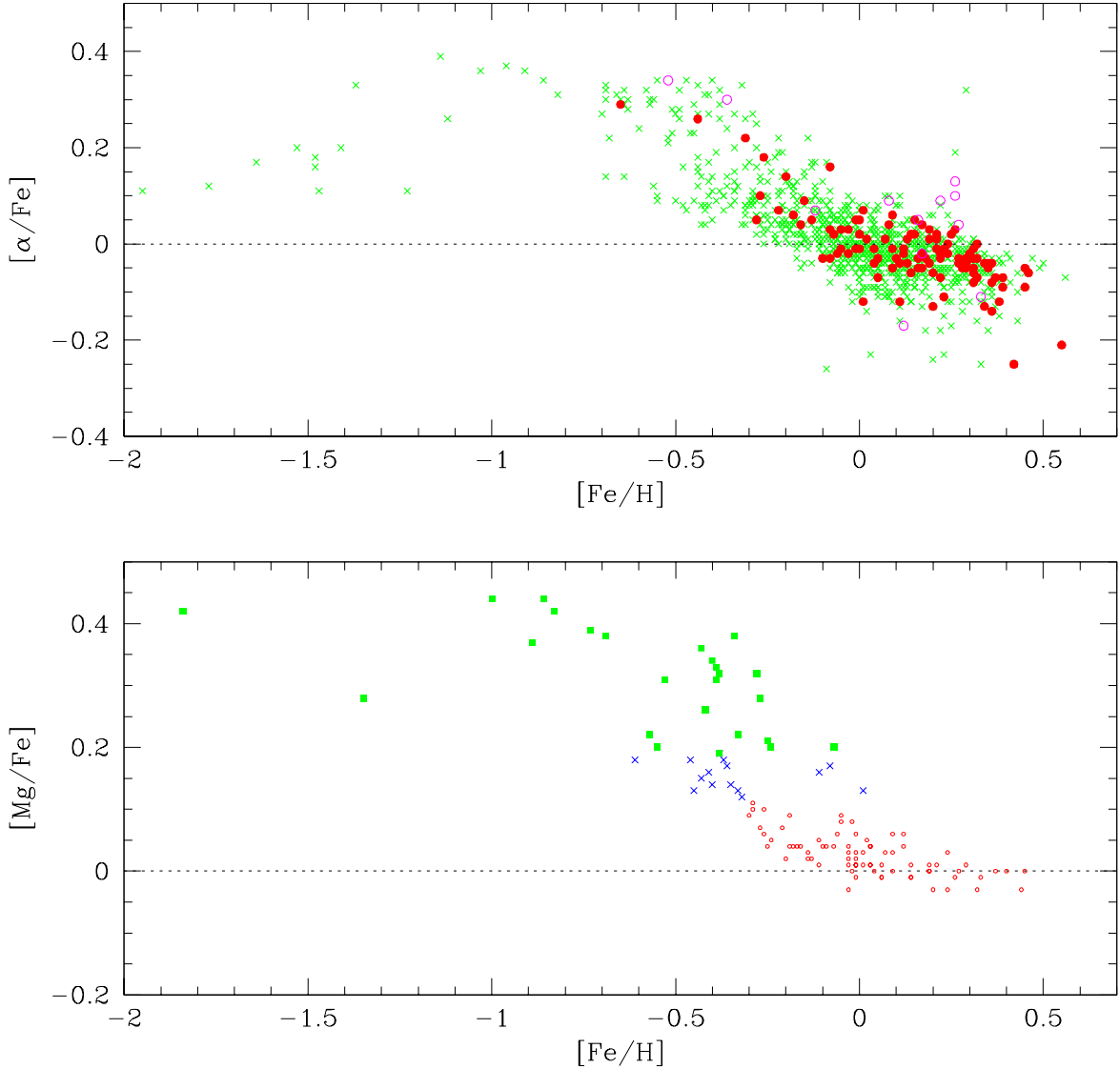}
\caption{ Thick disk planetary systems: The lower panel plots $\alpha$-element abundances (as exemplified by [Mg/Fe]) for nearby stars from Fuhrmann (1998), where the open squares are disk dwarfs, the solid squares mark stars identified as members of the thick disk, and four-point stars mark transition objects. The upper panel plots [Ti/Fe]/[Fe/H] data from Valenti \& Fischer's (2005) analysis of stars in the Berkeley/Carnegie planet survey (crosses), together with results from other high-resolution abundance analyses of exoplanet hosts. The solid points mark VF05 stars known to have planetary companions; the open circles plot data from Gilli \etall (2006), Santos \etall (2006) (both [Ti/Fe] abundances) and Ecuvillon \etall (2006) ([O/Fe abundances). The five exoplanet hosts with [$\alpha$/Fe]$>0.2$ are listed in Table 2 and discussed in the text.}
\end{figure}

\begin{figure}
\figurenum{15}
\plotone{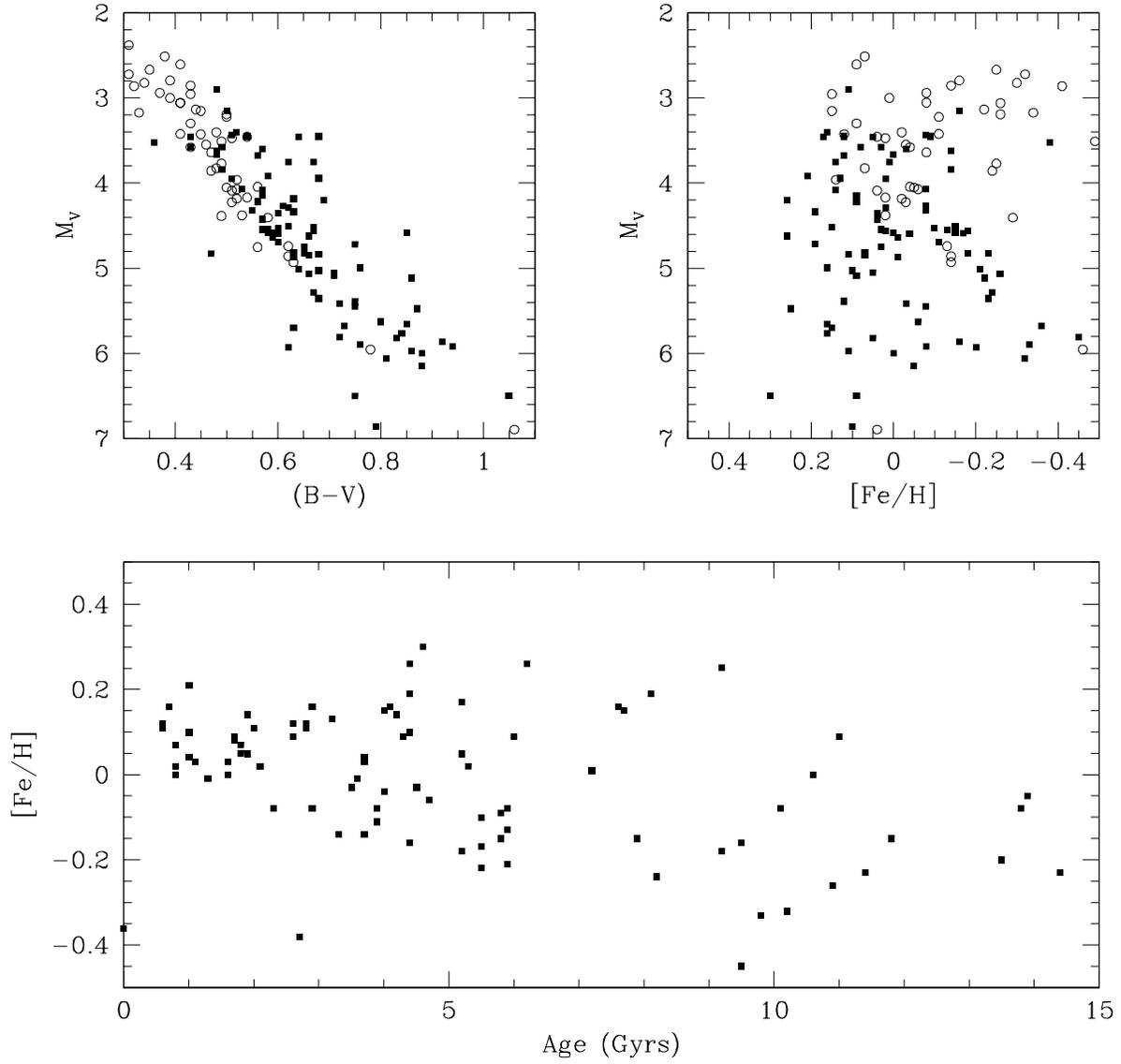}
\caption{ Colour-magnitude, M$_V$-metallicity and age-metallicity distributions for the 136 stars in Brown's high priority TPF target list. The 87 stars that are included in the VF05 dataset are plotted as solid squares. }
\end{figure}

\end{document}